\documentclass[12pt,english]{article}
\usepackage[T1]{fontenc}
\usepackage[latin9]{inputenc}
\usepackage{listings}
\usepackage{geometry}
\usepackage{float}
\usepackage{url}
\usepackage{amsthm}
\usepackage{amsmath}
\usepackage{amssymb}
\usepackage{graphicx}

\makeatletter

\providecommand{\tabularnewline}{\\}

 \theoremstyle{definition}
 \newtheorem*{defn*}{\protect\definitionname}

\usepackage{amsmath}
\usepackage{amssymb}
\usepackage{url}
\usepackage{listings}
\usepackage{url}
\@ifundefined{textcolor}{\usepackage{color}}{}

\makeatother

\usepackage{babel}
  \providecommand{\definitionname}{Definition}

\begin{document}

\title{A computer algebra user interface manifesto}

\author{David R. Stoutemyer%
\thanks{dstout at hawaii dot edu%
}}
\maketitle
\begin{abstract}
Many computer algebra systems have more than 1000 built-in functions,
making expertise difficult. Using mock dialog boxes, this article
describes a proposed interactive general-purpose wizard for organizing
optional transformations and allowing easy fine grain control over
the form of the result -- even by amateurs. This wizard integrates
ideas including:

$\bullet\;$flexible subexpression selection;

$\bullet\;$complete control over the ordering of variables and commutative
operands, with well-chosen defaults;

$\bullet\;$interleaving the choice of successively less main variables
with applicable function choices to provide detailed control without
incurring a combinatorial number of applicable alternatives at any
one level;

$\bullet\;$quick applicability tests to reduce the listing of inapplicable
transformations;

$\bullet\;$using an organizing principle to order the alternatives
in a helpful manner;

$\bullet\;$labeling quickly-computed alternatives in dialog boxes
with a preview of their results, using ellipsis elisions if necessary
or helpful;

$\bullet\;$allowing the user to retreat from a sequence of choices
to explore other branches of the tree of alternatives -- or to return
quickly to branches already visited;

$\bullet\;$allowing the user to accumulate more than one of the alternative
forms;

$\bullet\;$integrating direct manipulation into the wizard; and

$\bullet\;$supporting not only the usual input-result pair mode,
but also the useful alternative derivational and \emph{in situ} replacement
modes in a unified window.
\end{abstract}

\section{Introduction}

\begin{flushright}
``\textsl{Before the Lisp machine interface to Macsyma,}\\
\textsl{ computer algebra was like doing mathematics encumbered by
boxing gloves}.''\\
-- Bill Gosper
\par\end{flushright}

\noindent I am sorry Bill, but that user interface from 1988 \cite{Krausz}
disappeared with the Lisp machine, and its best features regrettably
have not yet been implemented in any of the current most powerful
computer algebra systems. Even the much earlier 1972 article \cite{BartonAndFitch}
discusses desirable features that are still missing from modern systems.

Many computer algebra systems have more than 1000 built-in functions.
Besides standard \textsl{mathematical functions} such as $\cos(\ldots)$
and classic higher transcendental functions, built-in functions often
include numerous \textsl{optional transformation functions} such as
$\mathrm{expand}(\ldots)$, $\mathrm{factor}(\ldots)$, $\mathrm{trigExpand}(\ldots)$,
$\mathrm{convert}(\ldots)$, and $\mathrm{simplify}(\ldots)$ that
supplement default simplification with various transformations. Besides
the expression being transformed, these transformational functions
often accept optional extra arguments such as a list of variables,
and/or various keyword arguments that control details such as the
amount of factoring. Moreover, most computer algebra systems have
numerous global \textsl{control variables} whose values help control
transformations done by these functions and/or by default.

Unlike $\cos(\ldots)$, the names and semantic details of these transformation
functions and control variables are not part of any standard mathematics
curriculum. Therefore it requires a long time to fully exploit such
systems well, and most users never do. Moreover, these names and behaviors
vary greatly between systems, making it challenging to become skilled
with more than one system in order to exploit their differing capabilities.
Consequently many users are frustrated because they don't know how
to make any system transform an expression to a desired form.

Worse yet, often there is \textsl{no} composition of functions and/or
or combination of control-variable settings capable of producing a
desired form. For example, users often want expansion with respect
to a certain proper subset of the variables, with polynomial coefficients
that are factored with respect to the other variables -- or want partial
fractions with respect to a certain proper subset of some variables,
with factored numerators and denominators.

This article address the usefulness of computer algebra systems as
productivity tools to help amateur users accomplish their tasks without
necessarily being aware of the underlying algorithms, transformation
functions and their nomenclature. This article presents some ideas
for enhancing the kind of wizard implemented on the Lisp Machine by:\vspace{-0.3em}

\begin{enumerate}
\item more flexible subexpression selection;\vspace{-0.3em}

\item complete control over the ordering of variables and commutative operands,
with well-chosen defaults;\vspace{-0.3em}

\item interleaving the choice of successively less main variables with applicable
function choices to provide detailed control without incurring a combinatorial
number of applicable alternatives at any one level;\vspace{-0.3em}

\item quick applicability tests to reduce the listing of inapplicable transformations;\vspace{-0.3em}

\item using an organizing principle to order the alternatives in a helpful
manner;\vspace{-0.3em}

\item labeling quickly-computed alternatives in dialog boxes with a preview
of the result, using ellipsis elisions if necessary or helpful;\vspace{-0.3em}

\item allowing the user to retreat from a sequence of choices to explore
other branches of the tree of alternatives -- or to return quickly
to branches already visited;\vspace{-0.3em}

\item allowing the user to accumulate more than one of the alternative forms;\vspace{-0.3em}

\item integrating direct manipulation into the wizard; and\vspace{-0.3em}

\item supporting not only the usual input-result mode, but also the useful
alternative derivational and \emph{in situ} replacement modes in a
unified window.\vspace{-0.3em}

\end{enumerate}
Section \ref{sec:Important-features} describes important features
to combine in a user interface. Section \ref{sec:Demos} presents
some mock examples of using the proposed wizard. Section \ref{sec:Design-issues}
addresses design issues and their resolution, with a summary in Section
\ref{sec:Summary}. The Appendix summarizes some important rational-expression
transformations that should be included in addition to those discussed
in subsection \ref{sub:CommonRationalTransformations}. The wizard
must also be aware of all of the many transformations specific to
irrational expressions that are built in or should be, but that is
an open-ended topic too large for discussion here. However, the wizard
should be implemented in an extensible way that allows users to add
components easily at run time for new transformations that they implement.

This article often uses ``Float'' as an abbreviation for ``floating-point
number''.

\section{Important features to combine in a user interface\label{sec:Important-features}}

Many of the ideas in this section have been implemented to some extent
in various computer algebra systems, but integrating them into a uniformly
designed user interface could greatly enhance the user experience
over that of any one current system.

Computer algebra often generates large expressions, and with current
RAM sizes measured in gigabytes, screen area is now the most precious
resource for most user's tasks.%
\footnote{The maximum number of characters simultaneously legible on multiple
high-resolution screens or sheets of paper is unimprovable.%
} Many of the ideas discussed here are concerned with attempting to
make the best use of that limited resource to help amateur and expert
users arrive quickly at the most comprehensible alternative output
forms.

\subsection{Enhancing the Lisp Machine Macsyma precedent}

Among the most helpful features of Lisp Machine Macsyma was that as
you moved the mouse over an expression, the minimal rectangle containing
a syntactically complete subexpression surrounding the mouse pointer
would automatically be framed, which is perhaps the entire expression.
A right click would open a drop-down menu of common transformations
such as factor and expand, or you could enter a function name of your
own. The selected transformation is applied to the framed subexpression.

This feature could be regarded as a \textsl{wizard} that helped users
quickly locate appropriate subexpressions for desired transformations,
then apply them to those subexpressions. This is important because
without subexpression selection and shortcuts for applying a desired
transformations to selected subexpressions, it is painful for even
expert users to force large expressions into anything near the form
they would prefer -- death by a thousand cuts and pastes.

Part of the pain is carefully reassembling a final result from independently
transformed subexpressions, then carefully deleting the distracting
debris of all the intermediate steps.

Even merely ordering commutative operands as desired is difficult
or impractical in most computer algebra systems. For example, it is
a constant irritant to be unable to transform a result such as $E=c^{2}m$
to $E=mc^{2}$ or, better yet, to prearrange that it will automatically
be ordered as desired.

\subsection{Qualitative analysis}

There is a computer algebra package that does \textsl{qualitative}
analysis, such as automatically determining whether an expression
is monotonic, convex, periodic, or has odd or even symmetry with respect
to each variable \cite{StoutemyerQualitativeAnalysis}. Such properties
are often of greater interest than any particular form of the expression.
Therefore the right click should also offer a \textsl{qualitative
properties} option.

\subsection{Including direct manipulation}

Direct manipulation provides a complementary way that major computer
algebra systems could make their user interfaces more helpful: With
Milo \cite{AvitzurMilo} or Theorist \cite{BonadioTheorisst} you
could use the mouse to select a term or a factor, then drag and drop
it to\vspace{-0.3em}

\begin{itemize}
\item reorder terms and factors,\vspace{-0.3em}

\item distribute a term over a factor,\vspace{-0.3em}

\item factor out a common factor from a sum of terms,\vspace{-0.3em}

\item transpose factors or terms from one side of an equation to the other.\vspace{-0.3em}

\end{itemize}
The selected subexpression could also be dropped into a variable in
another expression to substitute the subexpression for every instance
of the variable in that other expression. Also, expressions could
optionally be compressed by automatic or mouse-driven temporary replacement
of subexpressions with ellipses. For example, here is a temporal sequence
of Milo snapshots for dragging $x$ successively further right in
an equation \cite{KajlerAndSoifferSurvey}:
\[
\frac{\fbox{{x}}\mathrm{y}}{2}+\mathrm{x=1\quad\rightarrow\quad\frac{y\fbox{{x}}}{2}+x=1}\quad\rightarrow\quad\frac{\mathrm{y}}{2}\fbox{{x}}+\mathrm{x}=1\quad\rightarrow\quad\left(\frac{\mathrm{y}}{2}+1\right)\fbox{{x}}=1\quad\rightarrow\quad\frac{\mathrm{y}}{2}+1=\frac{1}{\mathrm{x}}
\]
Milo evolved to The Plotting Calculator, which is still available
and supported \cite{Avitzur}. The most recent version of Theorist
is named LiveMath\textsuperscript{tm}, which is also still available
and supported \cite{Theorist-1}. Both are oriented toward mathematics
through calculus, but direct manipulation should also be implemented
in other computer algebra systems, integrated with the transformation
wizard proposed here: After selecting subexpressions, one of the transformation
options, if applicable, should be ``drag and drop''.

\subsection{Collecting multiple alternatives}

The wizard generates and displays the results of alternative transformations
as the user explores a tree of successive applicable transformations.
The user can accumulate any number of these alternative results into
a list that is returned as the result if the user wants more than
one alternative. For example, as users interactively view alternative
factored and partial fraction forms for a rational expression, they
can indicate which ones they want included in a returned list of alternatives.
This is inspired by Wolfram|Alpha \cite{WolframAlpha}, which automatically
returns multiple alternative results. The difference here is that
the user can participate in a more thorough exploration of the possible
alternatives and select only those of interest.

\subsection{Input-result pairs \textit{versus} derivation steps \textit{versus}
replacement}

Most industrial-strength computer algebra systems use only the \textsl{input-result}
mode. This mode is particularly appropriate when the goal is to obtain
good final results in as few steps as is practical.

Some mathematics education programs such as Mathpert\textsuperscript{tm}
and the SMG application for some TI computer algebra products use
the \textsl{derivation} mode wherein the input is transformed to a
result by selecting successive subexpressions and choosing transformations
from menus, with the annotated result of each step displayed beginning
on a separate line. This mode is also often used in theorem proving
software \cite{TheryEtAl,UITP}. This mode is also good for expository
use by professionals when they want to explain in a presentation or
publication how a result is derived. For example, the multi-step derivational
style is used several times in this article.

Some programs such as The Graphing Calculator offer the \textsl{replacement}
mode wherein selected transformations replace the selected subexpressions
\emph{in situ}. This mode has the advantage of conserving screen space
by minimizing the amount of debris -- at the expense of not being
able to view the input and result simultaneously.

An industrial-strength computer-algebra system should offer all three
modes. The following observations can justify allowing mixtures of
all three modes in a single session window and name-space context:\vspace{-0.3em}

\begin{itemize}
\item An input-result pair can be regarded as a one-step instance of the
derivation mode.\vspace{-0.3em}

\item A one step \emph{in situ} replacement could be labeled with a two-button
setter bar such as\\
\includegraphics{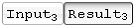} that toggles between the two.\vspace{-0.3em}

\item A multi-step \emph{in situ} replacement could have a slider bar between
these two endpoints, and perhaps also a Play button that does a slide
show or an animation.\vspace{-0.3em}

\item A right click could offer the option of changing previous computations
between these three modes, such as\vspace{-0.3em}

\begin{itemize}
\item collapsing a derivation sequence to an input-result pair or to an
\emph{in situ} replacement,\vspace{-0.1em}

\item expanding an input-result pair to a refinable derivation sequence
that was automatically used to create it.
\end{itemize}
\end{itemize}
The model-view-controller paradigm is a good way to achieve this multi-view
software design \cite{Model-view-controller}.

\subsection{\{Undo$^{m}$, redo$^{n}$\}}

Anyone who has used software with a well designed essentially unlimited
undo-redo capability knows how aggravating it is to return to software
that offers only one step of undo -- perhaps with no redo. With current
RAM capacities measured in gigabytes there is no excuse for this.
Internet browsing has familiarized users with using the ``$\leftarrow$''
and ``$\rightarrow$'' buttons together with a drop-down browse
history list to revisit easily throughout the tree of past web page
visitations. The wizard can use the same techniques and temporarily
save all recent closed dialog boxes for quick regeneration.

\subsection{Dynamically created dialog boxes specialized for the example}

The variety of mathematics examples is so great that general-purpose
dialog boxes created when a computer algebra system is built would
be unpleasantly cumbersome to use:\vspace{-0.3em}

\begin{itemize}
\item They would have numerous distracting grayed-out controls.\vspace{-0.3em}

\item They would entail numerous subsidiary dialog boxes to accommodate
all of the inapplicable entries without making each dialog box unreasonably
large.\vspace{-0.3em}

\item They would contain lengthy or awkward wording such as ``variable
or variables'' or ``variable(s)'' to correctly accommodate both
singular and plural cases without distracting grammatical errors.\vspace{-0.3em}

\end{itemize}
Thus custom dialog boxes specialized to the framed subexpression must
be created at run time.

\subsection{Adapt to the user's level and goals}

Computer algebra is being used by students from beginning secondary
school algebra through graduate-level mathematics -- and by professional
mathematicians, scientists, engineers, economists etc.\ at many different
levels of sophistication. The number of potential users declines rapidly
with increasing mathematics level. However, most computer algebra
systems are designed for the higher levels of this spectrum. Consequently
the most powerful general-purpose systems are quite daunting to most
potential users. For example, in many courses from secondary school
algebra through university real-variable calculus:\vspace{-0.3em}

\begin{enumerate}
\item Many students know nothing about hyperbolic functions, higher transcendental
functions and hypergeometric functions.\vspace{-0.3em}

\item Most students have not encountered many standard mathematics symbols
and notations such as $\exists$, $\forall$, $\neg$, $\vee$, $\wedge$,
$\Re$, $\Im$, $\mathbb{Z}$ and $\mathbb{Q}$.\vspace{-0.3em}

\item Most students know nothing about terminology such as algebraic groups,
rings, fields, ideals, varieties, and square-free factorization.\vspace{-0.3em}

\item The expression $1/0$ is usually or always regarded as undefined rather
than as $\pm\infty$ or a circle of infinite radius in the complex
plane.\vspace{-0.3em}

\item The expression $\sqrt{-1}$ is usually or always regarded as undefined
rather than $i$.\vspace{-0.3em}

\item The expression $(-1)^{1/3}$ is usually taken to mean $-1$ rather
than $1/2+i\sqrt{3}/2$.\vspace{-0.3em}

\end{enumerate}
The mathematically weakest students and professionals who could most
benefit from computer algebra are most intimidated by the appearance
of such unknown function names, symbols and nomenclature in their
dialog boxes and results. Often this intimidation and the consequent
loss of self esteem terminates receptivity to learning effective use
of the computer algebra system.

In computer aided instruction there are efforts to automatically infer
the level and overall goals of users, then adapt the interface accordingly.
Those techniques are not explored in this article, and the termination
of receptivity might occur before enough input occurs to make an accurate
inference. However, one easy way to accomplish many of the benefits
of such customization is for the first dialog box of a session to
have a button labeled ``Session preferences'' that opens a dialog
such as the following if pressed:

\noindent \begin{center}
\includegraphics[bb=0bp 0bp 578bp 264bp]{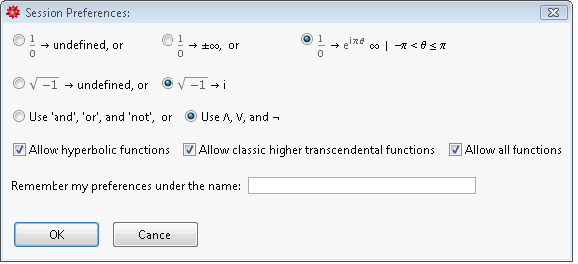}
\par\end{center}

\noindent The defaults should be those of the computer algebra system,
but the more elementary alternatives should appear first in each row
of alternatives to minimize alarming elementary users. 

Another complementary alternative to matching student mathematical
level is to enable the easy creation of named shell programs and their
icons that launch the computer algebra system then immediately set
appropriate preferences automatically. An instructor can then create
such shells named, for example, MapleForAlgebra1 or MathematicaForCalculus1.

\section{Examples of using the wizard\label{sec:Demos}}

This section contains examples of using the proposed wizard.

\subsection{Simplifying an ugly expression\label{sec:Demo}}

The \textsl{Mathematica} $\mathrm{CreateDialog}[\ldots]$ function
is a convenient way to create new dialog boxes at run time. Thus I
used $\mathrm{CreateDialog}[\ldots]$ to create dialog boxes that
are appropriate for specific examples, without bothering to attach
these boxes to each other or to any \textsl{Mathematica} transformation
functions. The mock intermediate and final results are not that of
any particular current computer algebra system, but rather what I
wish they would produce -- especially with regard to ordering of factors
and terms. For example, suppose that \textsl{either an input or a
result} of previous steps is
\begin{multline}
\boldsymbol{(}p^{2}qr^{3}+p^{2}qr^{2}s+p^{2}qr^{2}t+p^{2}qrst+p^{2}r^{4}+p^{2}r^{3}s+p^{2}r^{3}t+p^{2}r^{2}st+pqr^{4}+pqr^{3}s+pqr^{3}t+pqr^{2}st+\\
2pqrs+2pqrt+4pqr+4pqst+2pqs+2pqt+pr^{5}+pr^{4}s+pr^{4}t+pr^{3}st+pr^{2}s+2pqrs+\\
2pqrt+4pqr+4pqst+2pqs+2pqt+pr^{5}+pr^{4}s+pr^{4}t+pr^{3}st+pr^{2}s+pr^{2}t+2pr^{2}+\\
2prst+2prs+2prt+2pr+2pst+ps+pt+qr^{2}s+qr^{2}t+2qr^{2}+2qrst+2qrs+2qrt+\\
2qr+2qst+qs+qt+2r^{2}s+2r^{2}t+4r^{2}+4rst+2rs+2rt\boldsymbol{)/}\\
\boldsymbol{(}pqr^{2}+pqrs+pqrt+pqst+pr^{3}+pr^{2}s+pr^{2}t+prst+qr^{3}+qr^{2}s+qr^{2}t+qrst+r^{4}+r^{3}s+r^{3}t+r^{2}st\boldsymbol{)}.\label{eq:ExpandedOverExpanded}
\end{multline}

I have no \textsl{particular} goal form in mind, but I would like
a result that is more concise and comprehensible -- and more efficient
for substitution of numbers. I position the mouse pointer between
the left margin and the expression, thus framing the entire expression,
then right click and choose the ``qualitative properties'' option,
which determines that the function $f(p,q,r,s,t)$ defined by this
expression has the permutational symmetry $s\leftrightarrow t$. This
was determined by simplifying $f(p,q,r,s,t)-f(p,q,r,t,s)$ to 0. (None
of the differences for the other $5\times4-1=19$ transpositions of
two variables yielded a difference of 0.)

I then right click again and choose ``transform''.%
\footnote{Alternatively, I can choose ``Transform \ldots{}'' from the main
menu bar or click \framebox{\sf{Transform}} on the main toolbar.%
} This opens the following dialog box courteously positioned just above
the subexpression if the subexpression is low on the screen, or just
below the subexpression otherwise%
\footnote{I hate it when a dialog box initially covers the information I need
to respond to it! %
}:

\noindent \begin{center}
\includegraphics{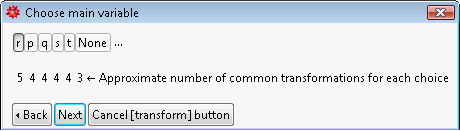}
\par\end{center}

Some transformations such as expansion of an improper ratio to a polynomial
plus a proper ratio require a designated variable. The \fbox{None}
button considers only transformations that do not require such a variable,
such as factoring or \textsl{polynomial} expansion with respect to
\textsl{all} variables. With that choice, the variables would be ordered
according to the analysis in \cite{Moses}.

The variables are listed in non-increasing order of an estimated number
of applicable common transformations because the choice of main variable
tends to have the greatest influence on the overall form of the result.
Choosing a main variable for which there are few alternative forms
tends to narrow the choices more than otherwise. For example, if we
chose $p$, $q$, $s$ or $t$ as the main variable, then there would
probably be fewer than 5 common alternatives for $r$ thereafter.
Thus button $r$ was initialized to \textsl{pressed} to encourage
lazy users such as me to accept it, which I do. This opens the following
dialog box:

\noindent \begin{center}
\includegraphics{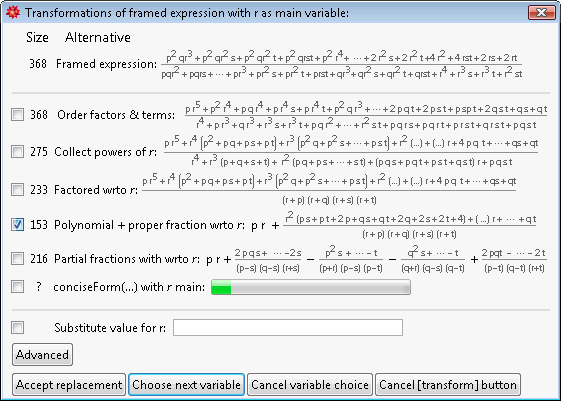}
\par\end{center}

\noindent \begin{flushright}
``\textsl{A lot of times, people don't know what they want until
you show it to them}.''\\
-- Steve Jobs
\par\end{flushright}

The \fbox{\sf{Advanced}} button lists alternatives that would interest
most users only occasionally for this example -- alternatives such
as continued fractions, Hornerization, expression in terms of Chebyshev
polynomials, or series approximations. 

The displayed \textsl{sizes} are some easily computed measure that
correlates approximately with the relative area that would be required
to display the entire alternative results. The initially checked boxes
are those having the smallest size.

The dialog box initially shows alternative results for all the alternatives
that can be computed in a \textsl{total} of at most 0.1 seconds --
with elisions if necessary to avoid scroll bars or using the entire
screen. User interface designers \cite{ShneidermanAndPlaisant} feel
that maximum acceptable response times are:\vspace{-0.3em}

\begin{itemize}
\item about 0.1 seconds for responses to a mouse click, key press, or anything
involving hand-eye coordination;\vspace{-0.3em}

\item about 1 second for opening a progress indicator, closing a dialog
box or reformatting a table;\vspace{-0.3em}

\item about 10 seconds for everything else, including displaying a graph
or completing an understandably time-consuming task. This ``mind
begins to wander'' threshold is not always achievable with computer
algebra.\vspace{-0.3em}

\end{itemize}
The \textsl{Mathematica} $\mathrm{Collect}[\ldots,r]$, $\mathrm{Factor}[\ldots]$,
$\mathrm{PolynomialQuotient}[\ldots,\ldots,r]$,\\
$\mathrm{PolynomialRemainder}[\ldots,\ldots,r]$ and $\mathrm{Apart}[\ldots,r]$
functions require a total of only 0.04 seconds on a dual-core 1.6
gigahertz computer to compute \textsl{Mathematica}-ordered versions
of the five initially-displayed results in this dialog box. Therefore
this dialog box could be created and displayed in an acceptable amount
of time.

The progress bar for the $\mathrm{conciseForm}(\ldots)$ alternative
appears within one second, indicating that it is still being computed
after the initial display of the dialog box. If that function doesn't
post progress messages, then a slug cyclically moves from left to
right to let the user know that the computer is working rather than
merely awaiting user input.%
\footnote{It is not yet customary to have computer algebra functions post progress
messages, but there are obvious candidate events for some algorithms.
For example, many algorithms for degree $n$ or for $n$ variables,
terms, factors, equations or columns process them one at a time, permitting
messages of the form ``$1/n$ \% done'', ``$2/n$ \% done'', ...
even if the time spent for each such step is likely to be rather uneven.
People are comforted by progress bars even when they are inaccurate.%
} The $\mathrm{conciseForm}(\ldots)$ alternative represents the system's
most powerful general-purpose simplification function such as the
\textsl{Mathematica} $\mathrm{FullSimplify}[\ldots]$ function, which
requires 2.4 seconds for this example. This is much less than the
time it requires for me to compare the five alternatives above $\mathrm{conciseForm}(\ldots)$
with the Framed expression. As such continuing computations complete,
their results are displayed in the dialog box -- partially elided
if necessary. They are listed below initially-presented results to
reduce displacement distraction when they complete. The $\mathrm{conciseForm}(\ldots)$
alternative completes with a size of 125, which is the smallest, so
that check box \textsl{also} becomes checked if I haven't already
pushed a button.

Every time a check box is checked that has not been checked before,
if the corresponding result is partially elided, then another dialog
box opens that merely displays the un-elided alternative result, with
scroll bars if necessary. For example, the initial ``Polynomial +
proper fraction wrt r'' choice opens the dialog box

\noindent \begin{center}
\includegraphics{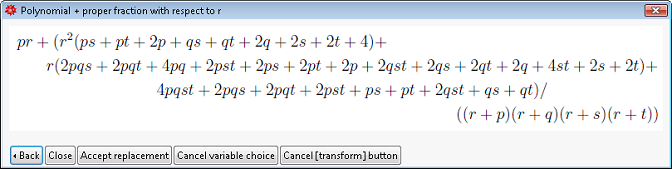}
\par\end{center}

\noindent and when $\mathrm{conciseForm}(\ldots)$ completes, it opens
the dialog box

\noindent \begin{center}
\includegraphics{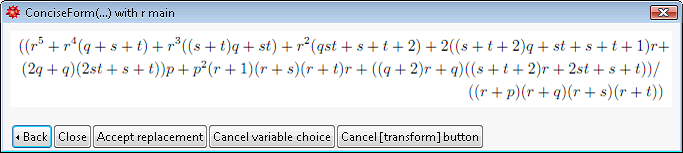}
\par\end{center}

Both alternatives are significant improvements over the original framed
expression (\ref{eq:ExpandedOverExpanded}), but both numerators are
still lengthy with no easily discerned pattern. The reasons for the
factored denominator in ``Polynomial + proper fraction'' are:\vspace{-0.3em}

\begin{itemize}
\item The factored denominator was already computed for the alternative
``Factored with respect to $r$''.\vspace{-0.3em}

\item The factored denominator is much more compact and informative than
the fully expanded original denominator.\vspace{-0.3em}

\item There is nothing in the phrase ``Polynomial + partial fraction''
that promises displaying the denominator expanded with respect to
$r$ that was used to compute the polynomial part and the numerator.\vspace{-0.3em}

\item It is easy to frame the factored denominator then expand it if desired.\vspace{-0.3em}

\end{itemize}
Although the ConciseForm result is slightly more compact, perhaps
I could improve the numerator of the proper fraction result because
I requested no more than a polynomial plus a proper fraction, and
the quickest path to that goal was to expend no extra effort on the
numerator beyond the collection with respect to the main variable
$r$ that was already done. Therefore in either the dialog box that
contains the elided or the complete version of the proper fraction,
I frame the entire numerator, right click, then choose ``transform''.
This recursively opens up a new dialog box to choose a main variable,
for which I again choose $r$ for consistency. The resulting displayed
alternatives include the factored numerator
\[
\left(r(p+q+2)+2pq+p+q\right)\left(r(s+t+2)+2st+s+t\right).
\]
This is much more compact, with insightful symmetries $p\leftrightarrow q$,
$s\leftrightarrow t$ and $\left[p,q\right]\leftrightarrow\left[s,t\right]$
that are also true of the denominator. Therefore I accept this replacement
in this \textsl{sub-problem} that is an alteration of the ``Polynomial
+ proper fraction'' alternative, thus transforming the expression
in that dialog box to\vspace{-0.2em}
\begin{equation}
pr+\dfrac{\left(r(p+q+2)+2pq+p+q\right)\left(r(s+t+2)+2st+s+t\right)}{(r+p)(r+q)(r+s)(r+t)}.\label{eq:PolyPlusFactoredProper}
\end{equation}
This is the nicest overall result so far, but before accepting it,
I notice that although every factor contains $r$, the first numerator
factor and the first two denominator factors are free of $s$ and
$t$, whereas the second numerator factor and the last two denominator
factors are free of $p$ and $q$. From experience I know that for
two ratios having disjoint variable sets or nearly so, common denominators
almost always increase bulk because there can be very little cancellation
in the resulting numerator. Thus conversely, partitioning the ratio
in (\ref{eq:PolyPlusFactoredProper}) into a ratio containing $\{r,p,q\}$
and a ratio containing $\{r,s,t\}$ then transforming each ratio to
partial fractions \textsl{might} reduce bulk. Consequently, I drag
the first numerator factor left of the ratio giving
\[
pr+\left(r(p+q+2)+2pq+p+q\right)\dfrac{\left(r(s+t+2)+2st+s+t\right)}{(r+p)(r+q)(r+s)(r+t)}.
\]
Then I drag the first two denominator factors under the former numerator
factor giving
\begin{equation}
pr+\left(\dfrac{r(p+q+2)+2pq+p+q}{(r+p)(r+q)}\right)\left(\dfrac{r(s+t+2)+2st+s+t}{(r+s)(r+t)}\right).\label{eq:PartitionedProperFractionFactors}
\end{equation}
(With the ability to select several non-adjacent subexpressions, I
could instead select in (\ref{eq:PolyPlusFactoredProper}) the first
factor of the numerator and the first two factors of the denominator,
then right click, then choose an action named something such as ``collect'',
``group'' or ``isolate'' -- or perhaps directly choose partial
fractions.)

Next I highlight the left ratio in (\ref{eq:PartitionedProperFractionFactors}),
choose $r$ as the main variable, then accept partial fraction expansion
with respect to $r$, then do similarly for the right factor, giving
\begin{equation}
pr+\left(\dfrac{p+1}{r+p}+\dfrac{q+1}{r+q}\right)\left(\dfrac{s+1}{r+s}+\dfrac{t+1}{r+t}\right).\label{eq:ProductOf2PartFracs}
\end{equation}
This is a very gratifying result compared to the equivalent input
(\ref{eq:ExpandedOverExpanded}), and I am happy with the ordering
of the terms and factors. Therefore I press the \fbox{\sf{Accept result}}
button, which opens the dialog

\noindent \begin{center}
\includegraphics{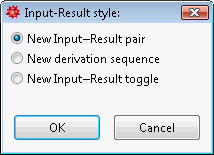}
\par\end{center}

If I choose ``New Input-Result pair'', then the main computer algebra
session window would be updated with a numbered input such as $\mathit{Input}_{4}:\;\mathrm{Transform}(\mathit{Result}_{3},\ldots)$
followed by expression (\ref{eq:ProductOf2PartFracs}) preceded by
a label such as $\mathit{Result}_{4}$. The purpose of the Input line
is to capture a programmatic way to transform $\mathit{Result}_{3}$
to $\mathit{Result}_{4}$ for purposes such as scripting. Most users
will not want to view the ugly details, but if I click on the ellipsis
in $\mathrm{Transform}(\mathit{Result}_{3},\ldots)$, then it expands
to a procedure that generates $\mathit{Result}_{4}$, such as\vspace{-0.4em}

\begin{minipage}[t]{1\columnwidth}%
\begin{lstlisting}
Block(Local(temp1,temp2,temp3,temp4),
  temp1 := PolynomialPlusProperFraction(Result3, r),
  temp2 := Factor(Numerator(Second(temp1)), r),
  temp3 := Factor(Denominator(Second(temp1)), r),
  First(temp1) +
    PartialFractions(First(temp2)/(First(temp3)*Second(temp3)), r)*
    PartialFractions(Second(temp2)/(Third(temp3)*Fourth(temp3)), r));
\end{lstlisting}
\end{minipage}

The ``New derivation sequence'' choice is similar, except it generates
a collapsible sequence of such pairs using labels such as $\mathrm{Intermediate}_{4,1}$,
$\mathrm{Intermediate}_{4,2}$, \ldots{}\,.

The ``New Input-Result toggle'' is similar to the ``New derivation
sequence'', except listing a single expression labeled with a two-button
setter bar.

\subsection{Transforming equations, inequalities and Boolean expressions}

The primary activity that users want to do with equations, inequalities
and systems thereof is to transform them into explicit solutions,
so why force users to learn numerous different function names with
different parameter semantics for solving different kinds of equations
and inequalities? If the user clicks the \fbox{Transform} button
when an entire equation, inequality or system thereof is framed, then
the wizard tries to return solutions. Here is an example for a differential
equation:

\noindent \begin{center}
\includegraphics{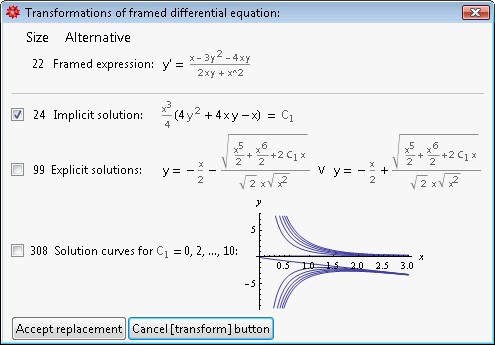}
\par\end{center}

\noindent Notice the synergy of multiple views of the solutions. When
the mouse pointer is over a curve, it is attached to a call-out displaying
the corresponding explicit solution containing the associated numeric
value for $C_{1}$. The plot range for $x$ and the set of values
for $C_{1}$ were chosen to insure that $y$ is real. Checking this
alternative opens a dialog with a magnified plot that gives the user
control over the plot ranges for $x$ and $y$ together with the set
of numeric values for $C_{1}$. The listed size of 308 is correlated
with the area of the plot in the session window if accepted.

Here is an example for a system of two nonlinear algebraic equations:

\vspace{-10em}

\noindent \begin{center}
\includegraphics[bb=0bp 0bp 660bp 419bp]{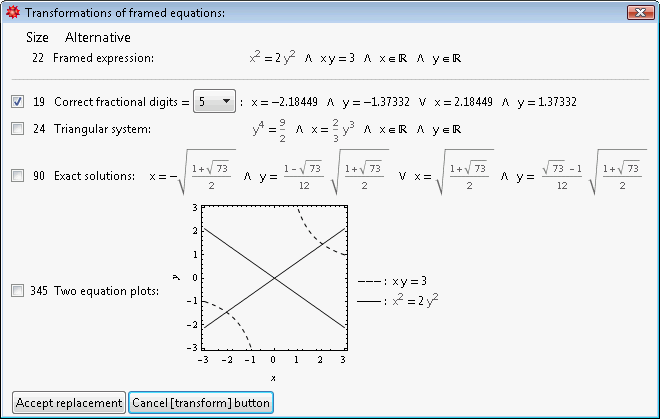}
\par\end{center}

\noindent \begin{center}
\vspace{-3em}

\par\end{center}
\begin{itemize}
\item \noindent The approximate solution was computed with \textsl{adaptive
precision interval arithmetic} to deliver guaranteed requested accuracy
initially set to correspond to six significant digits for nonzero
components. If interval arithmetic is inapplicable or is taking too
long for the specified accuracy, then the wizard tries \textsl{adaptive
significance arithmetic}. If that is also taking too long, the wizard
switches the popup digits setting to ``IEEE'' double and uses that.
The corresponding displayed phrases are ``Estimated correct fractional
digits'' or ``Approximate solutions of unknown accuracy'' respectively.\vspace{-0.3em}

\item \noindent The triangularized system is a reduced lexicographic Gröbner
basis, which might be the preferred alternative for parametrized systems
having exact explicit solutions that are messy or require unendurable
time to complete. If the system included inequalities, then there
would be a cylindrical algebraic decomposition instead.\vspace{-0.3em}

\item The $x$ and $y$ ranges in the plot were automatically set to include
a margin around the convex hull of all the isolated finite solutions
-- a margin small enough to resolve detail near the solutions but
large enough to provide useful context.\vspace{-0.3em}

\end{itemize}
Here is an example of transforming a Boolean expression:

\noindent \begin{center}
\includegraphics{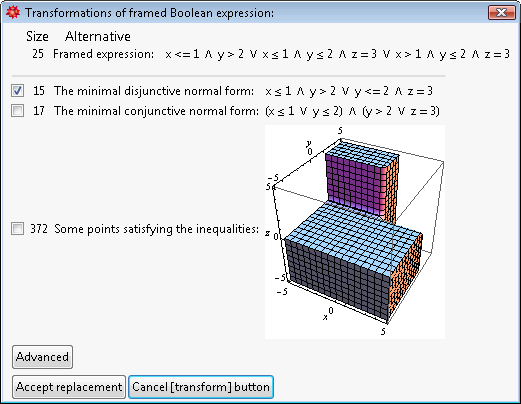}
\par\end{center}

\noindent The \fbox{\sf{Advanced}} button lists alternatives expressed
in terms of nands, nors, etc.

\subsection{The wizard is helpful even for mere numbers\label{sec:GoodForNumbersToo}}

Common useful internal representations for exact numbers are rational
numbers and irrational constant expressions such as $\sqrt{\pi}+\ln2$.
Common representations for approximate numbers are:\vspace{-0.3em}

\begin{enumerate}
\item software variable-precision Floats -- preferably with adaptive significance
tracking,\vspace{-0.3em}

\item IEEE double Floats to take advantage of fast hardware instructions;\vspace{-0.3em}

\item intervals whose endpoints are each independently an exact rational
number or a Float -- preferably adaptive precision for floating-point
endpoints, and allowing a disjoint union of such intervals with either
open or closed endpoints.\vspace{-0.3em}

\end{enumerate}
\textsl{Before} commencing a computation a user might want to transform
from an exact to an approximate representation to make the computation
faster -- or from an approximate to an exact representation to avoid
rounding errors. \textsl{After} a computation a user might want to
convert from an approximate to an exact representation to attempt
recovering an exact result -- or from an exact to an approximate representation
to make lengthy exact numbers more comprehensible or faster for purposes
such as plotting. Also, at the end of a computation a user might want
to simply alter the \textsl{display} of a number, such as displaying
an exact rational number factored or as an integer plus a proper fraction
or as a decimal fraction or in scientific form with a particular number
of fractional or significant digits. The wizard makes it easy to do
these transformations.

\subsubsection{Alternate forms for rational numbers}

If the framed subexpression is
\[
\fbox{{\ensuremath{\frac{1371742100137174210}{10973936901}}}},
\]
then the wizard could offer the dialog box

\noindent \begin{center}
\includegraphics{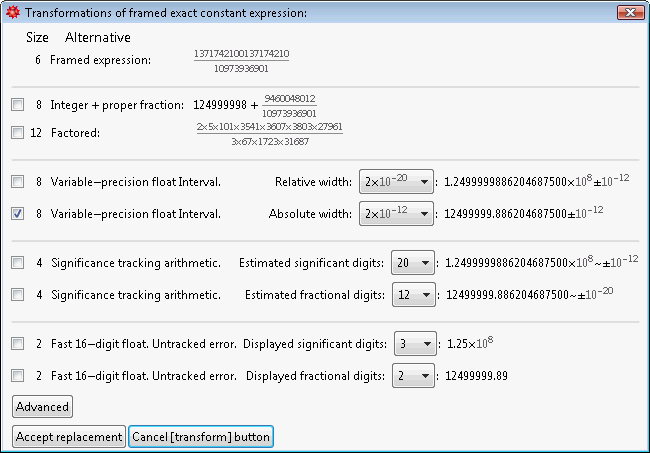}
\par\end{center}

\noindent In accordance with the recommendations of \cite{UsefulNumbers},
the approximate alternatives are ordered from intervals through bare
IEEE Floats to encourage more use of arithmetic that is closer in
spirit to exact arithmetic. The \fbox{\sf{Advanced}} button could
offer $p$-adic, continued fraction, and different radix representations.%
\footnote{In most systems, default simplification would immediately transform
a factored or continued fraction or integer plus proper fraction form
of a rational number back to a reduced ratio. Thus special passivity
is required to make such volatile forms appear in results, and such
non-idempotent forms are quite likely to disappear when such results
are used in subsequent inputs. For this reason, many systems return
a factored rational number as a list of pairs of bases and exponents,
etc. The wizard must correct for such impediments to direct substitution
of transformed subexpressions into the expression from which it came,
suppressing default simplification where necessary to preserve the
displayed overall result in standard mathematical form.%
}

Notice that this dialog provides useful supplementary information
about the framed number even if the user never intended to replace
the framed number: The user now has a good estimate for its magnitude,
can see that it is well approximated by $1.25\times10^{8}$, and that
both the numerator and denominator are composite but square free.

\subsubsection{Alternate forms for intervals and Floats}

If the framed subexpression is an interval, an IEEE double or a variable-precision
Float, then the wizard could offer alternatives including approximating
the number with an exact rational or irrational constant. For example,
if the framed number was the IEEE double $\fbox{{7.024814731040727}}$
or a reasonably close approximation to it, albeit perhaps \textsl{displayed}
with fewer significant digits, then the wizard could offer

\noindent \begin{center}
\includegraphics{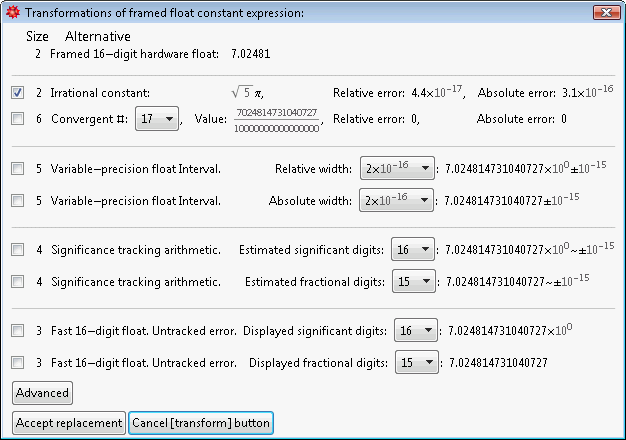}
\par\end{center}

Details matter. For example:\vspace{-0.3em}

\begin{itemize}
\item The alternate expressions are aligned, if practical, to make it easier
to compare them.\vspace{-0.3em}

\item If the framed Float displayed few digits, the initial displayed digits
for alternatives displays all or many digits -- and \emph{vice versa}.\vspace{-0.3em}

\end{itemize}
The delightful alternative $\sqrt{5}\,\pi$ was computed quickly by
the Maple $\mathrm{identify}(\ldots)$ function, for which there is
a more powerful free stand-alone version on the internet \cite{InverseSymbolicCalculator}.

\subsubsection{Alternate forms for non-real numbers}

For non-real numbers, which of \textsl{rectangular}, \textsl{unit
polar} and \textsl{exponential polar} form is most attractive depends
on the particular number. For example, compare

\[
\begin{array}{ccccc}
\mathit{Rectangular} &  & \mathit{Unit\: polar} & \rule[-9pt]{0pt}{26pt} & \mathit{Exponential\; polar}\\
\boldsymbol{7+5i} & = & \left(-1\right)^{\arctan\left(5/7\right)/\pi} & = & \sqrt{74}\, e^{\,\arctan\left(5/7\right)\, i},\\
-2\sin\left(\dfrac{3\pi}{14}\right)+2\cos\left(\dfrac{3\pi}{14}\right)i & = & \boldsymbol{2}\boldsymbol{\left(-1\right)^{5/7}} & = & 2e^{\,5\pi i/7},\\
2\cos\left(1\right)+2\sin\left(1\right)i & = & 2\left(-1\right)^{1/\pi} & = & \boldsymbol{2}\boldsymbol{e^{i}}.
\end{array}
\]
\textsl{Default} simplification would ideally \textsl{display} the
form that is most concise for each non-real number in a result even
if a different form is used internally. However, optional transformations
can conveniently offer all three alternatives.

\section{Design issues and their resolutions\label{sec:Design-issues}}

Challenging design issues include deciding:\vspace{-0.3em}

\begin{enumerate}
\item What set of transformations should the wizard consider?\vspace{-0.3em}

\item How can the wizard quickly estimate the number of applicable transformations
without knowing the next variable choice?\vspace{-0.3em}

\item When the next variable choice is known or \fbox{\sf{None}}, how can
the wizard quickly determine what subset of transformations are applicable
and schedule them so that a worthwhile number are completed quickly?\vspace{-0.3em}

\end{enumerate}
This section addresses these issues and a few others. However, there
are so many transformations that the wizard should know about that
this section concentrates on those that help resolve issues 2 and
3. For more completeness, the Appendix discusses additional transformations
that are relevant to the rational aspects of expressions. Transformations
of the \textsl{irrational} aspects of expressions is too large a topic
for treatment in this article.

First, three simple definitions:
\begin{defn*}
\textsl{Default simplification} is the result of pressing the \fbox{\sf{Enter}}
key or else perhaps \fbox{\sf{Shift}} \fbox{\sf{Enter}} with the
factory-default mode settings and no transformational or simplification
functions anywhere in the input expression.\vspace{-0.4em}

\end{defn*}
\noindent 
\begin{defn*}
A \textsl{functional form} is an expression of the form\vspace{-0.3em}
\[
f\left(\mathit{expression}_{1},\mathit{\, expression}_{2},\,\ldots\right)
\]
\vspace{-0.3em}
where $f$ is any function name.
\end{defn*}
\vspace{-0.4em}

\begin{defn*}
\begin{flushleft}
\textsl{Generalized variables} are the smallest subtrees of an expression
tree that are not a sum, difference, product, ratio, rational number,
Float, or reasonably regardable as an integer power.
\par\end{flushleft}
\end{defn*}
\vspace{-0.2em}
For example, $z$, $\pi$, $i$, $\cos(x+f(2))$, $z^{1/5}$, and
$3^{1/5}$ are generalized variables. In contrast, 3/4, $x/y$, and
$x-3$ are not. Also, $z^{2/5}$ and $3^{2/5}$ are not generalized
variables, because they can be regarded as $(z^{1/5})^{2}$ and $(3^{1/5})^{2}.$

The importance of generalized variables is that transformations that
are applicable with respect to variables in an expression can also
be applicable with respect to generalized variables in an expression.
For example, a user might want ordering, expansion or factoring with
respect to to $\pi$ and or $\cos(x+f(2))$. Additional transformations
might be applicable with respect to generalized variables that are
not merely indeterminates, such as \ $\cos(x+f(2))\rightarrow\sin(\pi/2-x-f(2))$,\\
$\cos(x+f(2))^{2}\rightarrow1-\sin(x+f(2))^{2}$, \ or\  $\pi\rightarrow3.14159$\,.

As a prerequisite to discussing the wizard, it is helpful to organize
the most important optional transformations offered by most computer
algebra systems into categories of related transformations. For simplicity,
the discussion addresses only constant ground domains that are common
scalar numeric domains of characteristic 0. However, much of the discussion
is relevant to other ground domains such as $\mathbb{Z}_{m}$ or $\{\mathrm{true},\,\mathrm{false}\}$.

\subsection{\label{sub:Different-forms-for-differentVars}Different transformations
for different generalized variables}

\begin{flushright}
``\textsl{To each his own}.''\\
-- Cicero\\
\medskip{}
``\textsl{I got different strokes for different folks}.''\\
-- Muhammad Ali
\par\end{flushright}

Often users want certain transformations such as expansion or factoring
only with respect to certain variables. For example:\vspace{-0.3em}

\begin{itemize}
\item To compute the integral of $\left(x^{5}+\left(c+1\right)^{999}x+1\right)^{2}$
with respect to $x$, it is helpful to expand with respect to $x$,
but foolish to expand with respect to $c$.\vspace{-0.2em}

\item To solve\vspace{-0.2em}
\[
\left(c^{999}-1\right)\left(z^{2}-1\right)=0\quad|\quad c^{999}\neq1
\]
it is helpful to factor with respect to $z$, but foolish to factor
with respect to $c$.\vspace{-0.2em}

\end{itemize}
In these cases we would prefer either concise or mere default simplification
with respect to $c$.

When the user requests successive transformation for successive variables,
we do not want to destroy transformations done for prior variables.
Consequently, requested transformations are automatically mapped into
the largest subexpressions that do not contain variables that have
already been treated. For example:\vspace{-0.2em}

\begin{enumerate}
\item If the alternative for \textsl{expanding} a framed expression with
respect to the chosen main variable $x$ is\vspace{-0.2em}
\[
\left(y^{2}+2y+1\right)x^{2}-y^{2}+2y-1,
\]
then \textsl{factoring} this alternative with respect to $y$ gives
$(y+1)^{2}x^{2}-(y-1)^{2}$ rather than
\[
\left((y+1)x+y-1\right)\left((y+1)x-y+1\right).
\]

\item If the alternative for \textsl{factoring} a framed expression with
respect to the chosen main variable $x$ is\vspace{-0.2em}
\[
(y-1)(y+1)\left((y-2)(y+2)x+(z+1)^{2}\right)\left(x+(2y+1)(2y-1)\right),
\]
then \textsl{expanding} this alternative with respect to $y$ gives
\[
(y^{2}-1)\left((y^{2}-4)x+(z+1)^{2}\right)\left(x+4y^{2}-1\right).
\]

\item If the alternative for \textsl{factoring} a framed expression with
respect to the chosen main variable $x$ is\vspace{-0.2em}
\[
\left(y^{2}-1\right)\left((y^{2}-4)x+(z+1)^{2}\right)\left(x+4y^{2}-1\right),
\]
then\textsl{ factoring} this alternative with respect to $y$ gives
\[
(y-1)(y+1)\left((y-2)(y+2)x+(z+1)^{2}\right)\left(x+(2y-1)(2y+1)\right).
\]
(Notice that the wizard factored not only the \textsl{coefficients}
of powers of $x$ with respect to $y$, including the coefficient
of the zeroth power of $x$, but also the top-level \textsl{content}
$y^{2}-1$, because none of these contain $x$.)
\item If the alternative for \textsl{expanding} a framed expression with
respect to the chosen main variable $x$ is\vspace{-0.2em}
\[
\left((z+1)^{9}y^{2}+y+3\right)x^{2}+(z+1)^{9}x+(y+1)(y-1),
\]
then \textsl{expanding} this alternative with respect to $y$ gives
two distinct alternatives: \textsl{distributed} form
\[
(z+1)^{9}x^{2}y^{2}+yx^{2}+3x^{2}+(z+1)^{9}x+y^{2}-1,
\]
and the often more concise \textsl{recursive} form
\[
\left((z+1)^{9}y^{2}+y+3\right)x^{2}+(z+1)^{9}x+y^{2}-1,
\]
both of which are offered to the user. Expansion of $(z+1)^{9}$ will
be offered if the user proceeds to that last remaining variable rather
than balking or accepting an alternative already displayed.\vspace{-0.2em}

\end{enumerate}
Now consider the input $\sin(x)\left(\cos(2y)+1\right)\cos(x)$. If
the user is allowed to choose trigonometric expansion of multiple
angles for the generalized variable $\cos(2y)$ but choose the opposite
transformation for $\sin(x)$ and $\cos(x)$, then this product can
transform to the particularly concise equivalent $\sin(2x)\cos(y)^{2}$
because $\cos(2y)\equiv2(\cos(y)^{2}-1)$ and $\sin(x)\cos(x)\rightarrow\sin(2x)/2$.

Thus for maximum flexibility:\vspace{-0.3em}

\begin{enumerate}
\item The user should be able to choose the order of generalized variables.\vspace{-0.3em}

\item The user should be able to choose separate transformations for each
generalized variable.\vspace{-0.3em}

\item The choices for each variable should include $\mathrm{conciseForm}(\ldots)$
and mere reordering with any associated default simplification when
such results differ from the framed subexpression.\vspace{-0.3em}

\item Where there is expansion with respect to two or more successive variables,
both distributed and recursive forms should be offered if they are
not identical.
\end{enumerate}

\subsection{Control over the order of generalized variables\label{sub:OrderingVariables}}

\begin{flushright}
``\textsl{Order is the shape upon which beauty depends}.''\\
-- Pearl S. Buck
\par\end{flushright}

Subsection \ref{sub:Different-forms-for-differentVars} discussed
how collection of similar powers of a generalized variable can be
recursively applied to the resulting collected coefficients to perform
transformations for successive generalized variables in any order.
However, current computer algebra systems would nonetheless impose
their built-in ordering rules to the resulting factors and terms.
Therefore the displayed ordering of factors in terms and of terms
in multinomials might not correspond to the recursive most main to
least main order in which the user has treated successive variables.
For example after requesting expansion with respect to $y$ with coefficients
that are factored with respect to all other variables a user might
obtain a result such as
\[
x^{3}+y^{3}(z+1)^{2}+y^{2}z
\]
rather than the more appropriate result
\[
(z+1)^{2}y^{3}+zy^{2}+x^{3}.
\]
Also, regardless of the requested order, output for Newton's definition
of force might be displayed as
\begin{equation}
f=am\label{eq:NewtonForce}
\end{equation}
which is visually quite disturbing despite its compliance with the
usual alphabetical ordering convention for variables in a monomial.
Consequently:\vspace{-0.6em}

\noindent \begin{center}
\framebox{Optional transformations should include control over the \textsl{displayed} ordering of factors and terms.}
\par\end{center}

The few systems that give such control tend to do so indirectly and
incompletely via control over the ordering of generalized variables.
For example:\vspace{-0.2em}

\begin{itemize}
\item In the Maxima computer algebra system the desired order $f=ma$ for
result (\ref{eq:NewtonForce}) can be accomplished by a declaration
such as $\mathrm{ordergreat}(a,m)$ or $\mathrm{orderless}(m,a)$.
If the user has issued an $\mathrm{ordergreat}(\ldots)$ and a non-conflicting
$\mathrm{orderless}(\ldots)$ declaration, then all other variables
order between the least of the great and the greatest of the least,
alphabetically. The effect is global from the time of a declaration
until all declared orders are deleted with an $\mathrm{unorder}()$
declaration, which \textsl{must} be used between any two $\mathrm{ordergreat}(\ldots)$
declarations or between any two $\mathrm{orderless}(\ldots)$ declarations.
Maxima also provides another mechanism for overriding default alphabetical
ordering:\vspace{-0.2em}
\[
\mathrm{declare}\left(\mathit{variable}_{1},\, property_{1},\,\mathit{variable}_{2},\, property_{2},\,\ldots\right)
\]
\vspace{-0.2em}
\negthinspace{}\negthinspace{}gives each variable the corresponding
property. Possible properties include \textsl{constant}, \textsl{scalar},
and \textsl{mainvar}. The command $\mathrm{remprop}(\mathit{variable},\, property)$
can be used to remove such a declaration. Using $\prec$ to represent
``less main'',\vspace{-0.3em}
\[
\mathit{constants}\prec\mathit{scalars}\prec\mathit{undeclared}\prec\mathit{mainvars}.
\]
\vspace{-0.3em}
\negthinspace{}\negthinspace{}By default, alphabetical order is
used within each of these categories.\vspace{-0.2em}

\item The Reduce computer algebra system has an order declaration that is
similar to $\mathrm{ordergreat}(\ldots)$, except that more than one
cumulative order declaration is allowed before a declaration\vspace{-0.3em}
\[
\mathrm{order\; nil};
\]
\vspace{-0.3em}
\negthinspace{}\negthinspace{}which clears all such ordering declarations.
The Reduce order declaration also accepts functional forms and built-in
literal constants such as $\pi$, which is important.\vspace{-0.2em}

\item These and some other systems provide some control over the ordering
of special distributed polynomials for Gröbner bases, but that is
not very helpful for controlling the ordering of factors and terms
in \textsl{general} expressions.\vspace{-0.2em}

\end{itemize}

\subsection{Common alternate forms for the \textsl{rational} aspect of expressions\label{sub:CommonRationalTransformations}}

Many computer algebra systems have separate functions for common denominators,
various factorization levels, and various levels of polynomial or
partial fractions expansions. This subsection describes how, for any
particular ordering of generalized variables, these traditionally
disparate concepts can be organized into a single topologically sorted
list of partially-ordered alternatives varying from the most complete
commonly-named factorization through the most complete commonly-named
expansion offered by many computer algebra systems. This \textsl{organizing
principle} greatly simplifies the transformation wizard by preventing
selection of a set of contradictory transformations and by making
the trade-off consequences in this list more obvious.

This subsection concerns only addition, subtraction, multiplication,
division and integer powers, but most of the ideas also apply recursively
to rational compositions of generalized variables and to fractional
powers. Moreover, this subsection discusses only factoring, common
denominators and expansion because they are most relevant to estimating
quickly how many transformations are applicable for each variable
and for quickly determining exactly which common transformations are
applicable for a particular variable. The Appendix discusses additional
rational transformations.

\subsubsection{Reasons for common denominators, factoring and expansion}
\begin{defn*}
A \textbf{candid expression} is one that is not equivalent to an expression
that visibly manifests a simpler expression class \cite{Stoutemyer10commandments}.

As counterexamples:\vspace{-0.3em}
\end{defn*}
\begin{itemize}
\item The expression $x(y+1)-xy$ is \textsl{not} candid because it contains
the superfluous variable $y$.\vspace{-0.3em}

\item The expression $(x+1)^{2}-x^{2}$ is \textsl{not} candid because it
appears to be quadratic but is actually linear.\vspace{-0.3em}

\item The expression $(x+1)/(x^{2}+2x+1)$ is \textsl{not} candid because
it is equivalent to $1/(x+1)$, which has lower numerator and denominator
degrees.
\end{itemize}
Reduction over a common denominator yields a candid form for rational
expressions, because the resulting form has no superfluous variables
and has maximum possible cancellation of poles with coincident zeros.
For most computer algebra systems, any amount of factoring \textsl{includes}
reduction over a common denominator.

\subsubsection{A \textsl{univariate} partially-ordered set of factorization and
expansion levels\label{sub:UnivariateFactorizationAndExpaqnsion}}
\begin{enumerate}
\item For univariate factoring there are names for certain amounts of exact
or approximate factoring based on multiplicities and the desired numeric
coefficient domain of the factors:\vspace{-0.4em}

\begin{enumerate}
\item \textsl{term primitive},%
\footnote{The \textsl{term content} of a univariate polynomial is the gcd of
the numeric coefficients times the smallest occurring power of the
variable. Factorization into the term content times the \textsl{term
primitive part} forces a common denominator if any coefficient has
a denominator, because with polynomials $A$, $B$, $C$ and $D$,\vspace{-0.3em}
\[
\frac{A}{B}+\frac{C}{D}\:\equiv\: AB^{-1}+CD^{-1}\quad\rightarrow\quad(AD+BC)B^{-1}D^{-1}\:\equiv\:\frac{AD+BC}{BD}\quad\rightarrow\quad\frac{\left(AD+BC\right)/G}{\left(BD\right)/G}
\]
\vspace{-0.3em}
\negthinspace{}\negthinspace{}where $G\,\leftarrow\,\gcd(AD+BC,\, BD)$.%
}\vspace{-0.2em}

\item \textsl{square free},\vspace{-0.2em}

\item \textsl{over the integer}s $\mathbb{Z}$,\vspace{-0.2em}

\item \textsl{over the Gaussian integers} $\mathbb{Z}[i]$,\vspace{-0.2em}

\item \textsl{over particular algebraic extensions},%
\footnote{As a convenience in \textsl{Mathematica,} $\mathrm{Factor}[\mathit{\mathit{expression},}\,\mathrm{Extension}\rightarrow\mathrm{Automatic]}$
automatically uses extensions implied by the complex unit $i$ and/or
any radicals \textsl{present} in $\mathit{expression}$. For example,\vspace{-0.3em}
\begin{eqnarray*}
\mathrm{Factor}[x^{2}+x-2+\sqrt{2},\,\mathrm{Extension}\rightarrow\mathrm{Automatic]} & \rightarrow & -\left(-x-1+\sqrt{2}\right)\cdot\left(x+\sqrt{2}\right)
\end{eqnarray*}
\vspace{-0.2em}
\negthinspace{}\negthinspace{}However, $\mathrm{Factor}[x^{2}+2\cdot\sqrt{2\cdot}x-1,\,\mathrm{Extension}\rightarrow\mathrm{Automatic]}\;\boldsymbol{\not\rightarrow}\;\left(x+\sqrt{2}-\sqrt{3}\right)\cdot\left(x+\sqrt{2}+\sqrt{3}\right)$
because $\sqrt{3}$ is not in the given polynomial. We must instead
use $\mathrm{Factor}[x^{2}+2\cdot\sqrt{2\cdot}x-1,\,\mathrm{Extension}\rightarrow\mathrm{\{\sqrt{2},\,\sqrt{3}\}]}$
to obtain this factorization, but how many users would know to include
$\sqrt{3}$\,?%
}\vspace{-0.2em}

\item \textsl{exact reasonably absolute},%
\footnote{This is what is usually expected of algebra through calculus students
for purposes such as solving equations or integrating rational functions:
Algebraic extensions implied by radicals in the input together with
use of the quadratic formula and $n$\textsuperscript{th} roots to
factor binomials.\textsl{ Derive} offers this option but also includes
cubic and quartic formulas, which tends to generate unreasonably messy
factorizations.%
}\vspace{-0.2em}

\item \textsl{exact absolute},%
\footnote{This means whatever algebraic extension is necessary to factor the
polynomial as much as possible, without the extension being provided
by the user. Reference \cite{fourLecturesAbsFactor} discusses some
algorithms for this. Some systems appear to use absolute factorization
in their functions that solve systems of polynomial equations and
integrate, but unfortunately appear not to offer it as a built-in
factorization option. Therefore many computer algebra systems cannot
directly factor $x^{2}+2x-1$ into $(x+1+\sqrt{2})(x+1-\sqrt{2})$
without assistance, which any beginning algebra student can do!.

Attempted exact absolute factorization might consume an intolerable
amount of computing time, or resulting factors might entail intolerably
messy nested radicals or intolerably messy subexpressions containing
functional forms named something such as Root. This is why I list
the \textsl{exact reasonably absolute} level of factorization.%
}\vspace{-0.2em}

\item \textsl{approximate absolute over the floating-point real numbers}
$\tilde{\mathbb{R}}$,\vspace{-0.2em}

\item \textsl{approximate absolute over the floating-point complex numbers}
$\tilde{\mathbb{C}}$.%
\footnote{Alternatives (h) and (i) are \textsl{approximations} rather than equivalence
transformations. Some methods for exact absolute factorization begin
from an approximate absolute factorization that is often preferable
to the resulting messy exact factorization!%
}\vspace{-0.2em}

\end{enumerate}
\item For systems that support variable precision Floats, users can choose
the precision level for alternatives (h) and (i). For systems that
offer significance and/or interval arithmetic, those alternatives
order immediately before (h) for real numbers and before (i) for non-real
numbers.\vspace{-0.2em}

\item \noindent With $F_{1}\succeq F_{2}$ denoting the fact that for a
given example, a factorization at level $F_{1}$ is either identical
to a factorization at level $F_{2}$ or is a further splitting of
the factorization at level $F_{2}$, we have\vspace{-0.2em}
\[
\begin{array}{c}
\mathrm{\mathbb{Z}[\mathit{i}]\:\succeq\:\mathbb{Z}\:\succeq\: square\: free\:\succeq\: term\: primitive},\rule[-0.2em]{0em}{1.2em}\\
\mathrm{exact\: absolute}\:\succeq\:\mathrm{specific\: algebraic\: extensions}\:\succeq\:\mathbb{Z},\rule[-0.2em]{0em}{1.4em}\\
\mathrm{exact\: absolute}\:\succeq\:\mathrm{reasonably\: exact\: absolute}\:\succeq\:\mathbb{Z},\rule[-0.2em]{0em}{1.4em}\\
\tilde{\mathbb{C}}\:\succeq\:\tilde{\mathbb{R}}\:\succeq\:\mathbb{Z},\rule[-0.2em]{0em}{1.3em}\\
\mathbb{\tilde{\mathbb{C}}\:\succeq\: Z}[i]\rule[-0.2em]{0em}{1.3em}.
\end{array}
\]
\vspace{-0.2em}
\negthinspace{}\negthinspace{}\negthinspace{}Thus these factorization
levels form a \textsl{directed acyclic graph.} that we can topologically
sort into one of several alternative lists, such as order 1(a) through
1(i) above.%
\footnote{Actually, different specific algebraic extensions generally form a
directed acyclic subgraph because, for example, we could have any
one, two or all three of the extensions $\sqrt{2}$, $\sqrt{3}$ and
$\sqrt{5}$, giving more than one path to $\{\sqrt{2},\,\sqrt{3},\,\sqrt{5}\}.$
These directed acyclic subgraphs are the field extension lattices
of Galois theory.%
}\vspace{-0.2em}

\item If we fully expand the product of the numerator factors and the product
of the denominator factors of a reduced ratio, then we have the reduced
ratio of two fully expanded polynomials. Despite the common denominator,
the result is an expanded polynomial when this reduced denominator
is 1 or when both the numerator and denominator are numeric. Therefore
this form is on the borderline between factored and expanded.\vspace{-0.2em}

\item The computer algebra built-into Texas Instruments hand held, Windows
and Macintosh products has a function $\mathrm{propFrac}\,(\mathit{expression},\,\mathit{variable})$
that expands $\mathit{expresssion}$ into a expanded polynomial with
respect to $\mathit{variable}$ plus a reduced ratio of two polynomials
that is \textsl{proper} with respect to $variable$. The $\mathrm{propFrac}\,(\ldots)$
function can easily be implemented using a polynomial quotient and
remainder function, and the resulting form is an appropriate next
node in our partial ordering from most factored to most expanded.
This form is often more concise than either a common denominator or
a partial fraction expansion. For example, this form was a key intermediate
step in the example of subsection \ref{sec:Demo}. For canonicality:\vspace{-0.4em}

\begin{enumerate}
\item The coefficients of the resulting expanded univariate polynomial part
that are not complex Floats can be normalized to Gaussian rationals
$\mathbb{Q}[i]$ or rationalized algebraic numbers.\vspace{-0.2em}

\item The denominator of the proper ratio can be made unit normal as described
in \cite{StoutemyerUnitNormal}.\vspace{-0.2em}

\item The numeric coefficients in the resulting proper ratio that are not
complex Floats can be normalized to Gaussian integers or algebraic
integers.\vspace{-0.2em}

\item If all of the denominator numeric coefficients are complex Floats,
then we can normalize their magnitudes -- such as making the largest
of the real and imaginary magnitudes in the denominator coefficients
be 1.0.\vspace{-0.3em}

\end{enumerate}
\item Polynomial expansion can be regarded as a special case of $\mathrm{propFrac}(\ldots)$
for when the denominator of the given reduced ratio is numeric --
perhaps 1.\vspace{-0.2em}

\item If the reduced ratio of two polynomials has a non-numeric denominator,
then the relevant adjective phrases 1(a) through 1(i) above can be
used to label successive nodes corresponding to the amount of denominator
factorization for corresponding partial fraction expansions.\vspace{-0.2em}

\item However, the square-free aspect of partial fraction expansion has
two variants in the partial ordering. In non-decreasing order of the
amount of expansion, adjective phrases applicable to the square free
aspect are:\vspace{-0.42em}

\begin{enumerate}
\item \textsl{incomplete}, meaning multiples of all powers of the same square-free
denominator factor are combined over a common denominator, and\vspace{-0.2em}

\item \textsl{complete}, meaning instead that for each resulting \textsl{very
proper} ratio $N(x)/D(x)^{m}$ with expansion variable $x$, $\deg_{x}(N(x))<\deg_{x}(D(x))$.%
\footnote{In contrast, for the \textsl{incomplete} square-free partial fraction
expansion we can guarantee only that $\deg_{x}(N(x))<\deg_{x}(D(x)^{m})$.
Many systems offer only complete expansions, but incomplete expansions
are adequate for most purposes and are often more concise!%
}\vspace{-0.2em}

\end{enumerate}
\end{enumerate}
Whenever a resulting numerator has more than one term, we can distribute
an associated denominator over the numerator terms. It is generally
unwise to distribute a multinomial denominator over the numerator
terms for purposes such as integration, and it almost always increases
bulk. However, it is helpful to do such a distribution for purposes
such as fragmenting a ratio into the greatest number of simplest possible
pieces for angle sum expansion.%
\footnote{If you must distribute multinomial denominators over numerator terms,
it is most efficient to wait until after the expansion is complete
in other respects.%
} The wizard can display both alternatives when they are not identical.

Table \ref{Flo:TableFactoredThruExpanded} shows the named alternative
forms for a univariate example of the reduced ratio of two expanded
polynomials.\vspace{-0.3em}

\begin{enumerate}
\item The first two rows and last two rows are approximations to all of
the other rows, which are equivalent to each other.\vspace{-0.3em}

\item The double line separating ``ratio of expanded polynomials'' and
``polynomial + proper ratio'' separates factored from expanded forms.\vspace{-0.3em}

\item Factors that differ from those of the preceding row are boldface.\vspace{-0.3em}

\item Wherever there is a sum in a numerator, the corresponding denominator
can optionally be distributed over the terms of the numerator.\vspace{-0.3em}

\item For each of these named levels an example can be constructed where
it is more concise than all of the other named levels. Therefore all
of the levels are important.%
\footnote{There are also unnamed intermediate levels such as splitting some
but not all of the square-free factors over $\mathbb{Z}$.%
}\vspace{-0.3em}

\item If a system doesn't offer built-in support for all of these named
levels and a wizard implementer is not inclined to add such support,
then:\vspace{-0.3em}

\begin{enumerate}
\item Missing intermediate factorization levels can be provided by over-factoring
then expanding appropriate subsets of factors.
\item Missing expansion levels can be provided by over expanding then combining
appropriate subsets of summands.\vspace{-0.3em}

\end{enumerate}
\item The input could be any of these expressions or any rational expression
that is equivalent to one of these expressions. If an input contains
Floats, then float-free alternatives can be obtained by using a function
such as the Maple $\mathrm{identify}(\ldots)$ function to determine
close rational or irrational constants \cite{InverseSymbolicCalculator}.
\end{enumerate}
\begin{table}[H]
\noindent \centering{}\caption{A univariate expression partially ordered from most factored to expanded }
\label{Flo:TableFactoredThruExpanded}%
\begin{tabular}{|c|c|}
\hline 
Amount of factor or expand & Boldface parts are different from the alternative above them\tabularnewline
\hline 
\hline 
$\tilde{\mathbb{C}}$ & $\frac{1.5\left(z-1.37\right)(z+5.7)(z-1.07+0.76i)(z-1.07+0.76i)\cdots\left(z+1.09+0.18i\right)\left(z+1.09-0.18i\right)}{z\left(z-1\right)^{2}\left(z+i\right)\left(z-i\right)\left(z+1.414\right)\left(z-1.414\right)\left(z+1.13\right)\left(z-1.04+0.82i\right)(z-1.04-0.82i)\cdots}_{\,}$\rule[-0.3em]{0em}{2.0em}\tabularnewline
\hline 
$\tilde{\mathbb{R}}$ & $\frac{1.5\cdots\boldsymbol{(z^{2}-2.13z+1.71)(z^{2}-1.13z+0.9)(z^{2}-0.05z+0.24)\cdots(z^{2}+2.19z+1.23)}}{z\left(z-1\right)^{2}\boldsymbol{(z^{2}+1)}\left(z+1.414\right)\left(z-1.414\right)\left(z+1.13\right)\boldsymbol{(z^{2}-2.08z+1.76)(z^{2}+0.95z+1.5)}}_{\,}$\rule[-0.3em]{0em}{2.0em}\tabularnewline
\hline 
 & \vspace{-1.1em}
\tabularnewline
\hline 
reasonably absolute ($\mathbb{Z}\left[i,\sqrt{2}\,\right]$) & $\frac{\boldsymbol{3(z^{12}+4z^{11}-9z^{10}+4z^{9}+z^{8}+13z^{6}-29z^{5}+7z^{4}+13z^{3}-25z^{2}+6z-6)}}{\boldsymbol{2}z\left(z-1\right)^{2}\boldsymbol{\left(z+i\right)\left(z-i\right)}\boldsymbol{(z+\sqrt{2})(z-\sqrt{2})(z^{5}+z+3)}}_{\,}$\rule[-0.3em]{0em}{2.0em}\tabularnewline
\hline 
$\mathbb{Z}[i]$ & $\frac{3(z^{12}+4z^{11}-9z^{10}+4z^{9}+z^{8}+13z^{6}-29z^{5}+7z^{4}+13z^{3}-25z^{2}+6z-6)}{2z\left(z-1\right)^{2}\left(z+i\right)\left(z-i\right)\boldsymbol{(z^{2}-2)}(z^{5}+z+3)}_{\,}$\rule[-0.3em]{0em}{2.0em}\tabularnewline
\hline 
$\mathbb{Z}$ & $\frac{3(z^{12}+4z^{11}-9z^{10}+4z^{9}+z^{8}+13z^{6}-29z^{5}+7z^{4}+13z^{3}-25z^{2}+6z-6)}{2z\left(z-1\right)^{2}\boldsymbol{(z^{2}+1)}(z^{2}-2)(z^{5}+z+3)}_{\,}$\rule[-0.3em]{0em}{2.0em}\tabularnewline
\hline 
square free & $\frac{3(z^{12}+4z^{11}-9z^{10}+4z^{9}+z^{8}+13z^{6}-29z^{5}+7z^{4}+13z^{3}-25z^{2}+6z-6)}{2z\left(z-1\right)^{2}\boldsymbol{(z^{9}-z^{7}-z^{5}+3z^{4}-z^{3}-3z^{2}-2z-6)}}_{\,}$\rule[-0.3em]{0em}{2.0em}\tabularnewline
\hline 
\textsl{term} primitive & $\frac{3(z^{12}+4z^{11}-9z^{10}+4z^{9}+z^{8}+13z^{6}-29z^{5}+7z^{4}+13z^{3}-25z^{2}+6z-6)}{2z\boldsymbol{(z^{11}-2z^{10}+2z^{8}-2z^{7}+5z^{6}-8z^{5}+2z^{4}+3z^{3}-5z^{2}+10z-6)}}$\rule[-0.3em]{0em}{2.0em}\tabularnewline
\hline 
ratio of expanded polynomials & $\frac{\boldsymbol{3z^{12}+12z^{11}-27z^{10}+12z^{9}+3z^{8}+39z^{6}-87z^{5}+21z^{4}+39z^{3}-75z^{2}+18z-18}}{\boldsymbol{2z^{12}-4z^{11}+4z^{9}-4z^{8}+10z^{7}-16z^{6}+4z^{5}+6z^{4}-10z^{3}+20z^{2}-12z}}_{\,}$\rule[-0.3em]{0em}{2.0em}\tabularnewline
\hline 
 & \vspace{-1.1em}
\tabularnewline
\hline 
polynomial + proper ratio & $\!\frac{3}{2}\!+\!\frac{\boldsymbol{18z^{11}-27z^{10}+6z^{9}+9z^{8}-15z^{7}+63z^{6}-93z^{5}+12z^{4}+54z^{3}-105z^{2}+36z-18}}{2z^{12}-4z^{11}+4z^{9}-4z^{8}+10z^{7}-16z^{6}+4z^{5}+6z^{4}-10z^{3}+20z^{2}-12z}_{\,}\negthinspace$\rule[-0.3em]{0em}{2.0em}\tabularnewline
\hline 
term primitive partial fraction & $\frac{3}{2}+\boldsymbol{\frac{3}{2z}+\frac{15z^{10}-21z^{9}+6z^{8}+3z^{7}-9z^{6}+48z^{5}-69z^{4}+6z^{3}+45z^{2}-90z+6}{2z^{11}-4z^{10}+4z^{8}-4z^{7}+10z^{6}-16z^{5}+44^{5}+6z^{3}-10z^{2}+20z-12}}_{\,}$\rule[-0.3em]{0em}{2.0em}\tabularnewline
\hline 
\negthinspace{}incomplete square free part frac\negthinspace{} & $\frac{3}{2}+\frac{3}{2z}+\boldsymbol{\frac{3z+3}{(z-1)^{2}}+\frac{12z^{8}-3z^{6}-3z^{4}+36z^{3}-18z+24}{2z^{9}-2z^{7}-2z^{5}+6z^{4}-2z^{3}-6z^{2}-4z-12}}_{\,}$\rule[-0.3em]{0em}{2.0em}\tabularnewline
\hline 
complete square free part frac & $\frac{3}{2}+\frac{3}{2z}+\boldsymbol{\frac{3}{(z-1)^{2}}+\frac{3}{2z-2}}+\frac{12z^{8}-3z^{6}-3z^{4}+36z^{3}-18z+24}{2z^{9}-2z^{7}-2z^{5}+6z^{4}-2z^{3}-6z^{2}-4z-12}_{\,}$\rule[-0.3em]{0em}{2.0em}\tabularnewline
\hline 
partial fractions over $\mathbb{Z}$ & $\frac{3}{2}+\frac{3}{2z}+\frac{3}{(z-1)^{2}}+\frac{3}{2z-2}+\boldsymbol{\frac{3z}{z^{2}-2}+\frac{3z}{z^{2}+1}}\boldsymbol{+\frac{3z^{2}-12}{2z^{5}+2z+6}_{\,}}$\rule[-0.3em]{0em}{2.0em}\tabularnewline
\hline 
partial fractions over $\mathbb{Z}[i]$ & $\frac{3}{2}+\frac{3}{2z}+\frac{3}{(z-1)^{2}}+\frac{3}{2z-2}+\frac{3z}{z^{2}-2}+\boldsymbol{\frac{3z}{2z+2i}+\frac{3}{2z-2i}}+\frac{3z^{2}-12}{2z^{5}+2z+6}_{\,}$\rule[-0.3em]{0em}{2.0em}\tabularnewline
\hline 
absolute partial fractions & $\frac{3}{2}\!+\!\frac{3}{2z}\!+\!\frac{3}{(z-1)^{2}}\!+\!\frac{3}{2z-2}\!+\!\boldsymbol{\frac{3}{2z+2\sqrt{2}}\!+\!\frac{3}{2z-2\sqrt{2}}}\!+\!\frac{3z}{2z+2i}\!+\!\frac{3}{2z-2i}\!+\!\frac{3z^{2}-12}{2z^{5}+2z+6}_{\,}$\rule[-0.3em]{0em}{2.0em}\tabularnewline
\hline 
 & \vspace{-1.1em}
\tabularnewline
\hline 
partial fraction over $\tilde{\mathbb{R}}$ & $\cdots\!+\!\boldsymbol{\frac{1.5}{z+1.41}\!+\!\frac{1.5}{z-1.41}\!+\!\frac{3.0z}{z^{2}+1}\!+\!\frac{0.162}{z+1.1}\!-\!\frac{0.18z-0.27}{z^{2}-2.1z+1.8}\!+\!\frac{0.018z+0.31}{z^{2}+0.95z+1.5}}_{\,}$\rule[-0.3em]{0em}{2.0em}\tabularnewline
\hline 
partial fraction over $\tilde{\mathbb{C}}$ & $1.5\!+\!\cdots\!-\!\boldsymbol{\frac{0.089+0.049i}{z-1.04+0.82i}\!-\!\frac{0.089-0.049i}{z-1.04-0.82i}\!+\!}\boldsymbol{\frac{0.0088+0.13i}{z+0.48+1.13i}\!+\!\frac{0.0088-0.13i}{z+0.48-1.13i}}_{\,}$\rule[-0.3em]{0em}{2.0em}\tabularnewline
\hline 
\end{tabular}
\end{table}

\subsubsection{\textsl{Multivariate} partially-ordered sets of factorization and
expansion levels\label{sub:MultivariateFactorExpand}}

References \cite{RaichevMultivariatePartialFractions,StoutemyerPartialFractions}
describes how to do multivariate partial fraction expansion with respect
to two or more successive generalized variables. As a degenerate case,
the expansion is \textsl{polynomial} expansion with respect to variables
that don't occur in the reduced common denominator of the given expression.

As shown that article, the number of terms in a multivariate partial
fraction expansion can depend on the ordering of the expansion variables.
Table \ref{Flo:BivariateFactorExpand} shows eight alternatives for
factoring and or expanding a bivariate example over $\mathbb{Z}$.
Notice how the partial fraction expansion with respect to $y$ in
the last two rows introduces poles at $x=\pm1$ into some individual
ratios, forcing the use of a piecewise result to avoid contracting
the domain of definition. Making a ratio proper can also cause this.
(Unfortunately, most current computer algebra systems quietly do such
domain reductions.) Common denominators remove these additively removable
singularities. 
\begin{table}[h]

\noindent \centering{}\caption{Some alternative recursive form factorizations and expansions over
$\mathbb{Z}$. \protect \\
f = \textbf{f}actored. e = \textbf{e}xpanded to incomplete partial
fractions.}
\label{Flo:BivariateFactorExpand}%
\begin{tabular}{|c|c|c|}
\hline 
\negthinspace{}\negthinspace{}\negthinspace{}\negthinspace{}1\textsuperscript{st}\negthinspace{}\negthinspace{}\negthinspace{}\negthinspace{} & \negthinspace{}\negthinspace{}\negthinspace{}2\textsuperscript{nd}\negthinspace{}\negthinspace{}\negthinspace{}\negthinspace{} & Result\tabularnewline
\hline 
\hline 
\negthinspace{}\negthinspace{}$\mathrm{f}_{x}$\negthinspace{}\negthinspace{} & \negthinspace{}\negthinspace{}$\mathrm{f}_{y}$\negthinspace{}\negthinspace{} & $\frac{(y-1)(y+1)(y^{2}+3)x^{3}-y(y+1)(y^{4}-y^{3}+y^{2}-3y-4)x-2y^{2}(y-1)(y^{2}+2)}{(x-y)(x+y)(y-1)(y+1)(y^{2}+2)}_{\,}$\rule[-0.7em]{0em}{2.1em}\tabularnewline
\hline 
\negthinspace{}\negthinspace{}$\mathrm{f}_{x}$\negthinspace{}\negthinspace{} & \negthinspace{}\negthinspace{}\negthinspace{}$\mathrm{e}_{y}$\negthinspace{}\negthinspace{}\negthinspace{} & $\frac{(y^{4}+2y^{2}-3)x^{3}-(y^{6}-2y^{3}-7y^{2}-4y)x-(2y^{5}-2y^{4}+4y^{3}-4y^{2})}{(x-y)(x+y)(y^{4}+y^{2}-2)}_{\,}$\rule[-0.7em]{0em}{2.1em}\tabularnewline
\hline 
\negthinspace{}\negthinspace{}$\mathrm{f}_{y}$\negthinspace{}\negthinspace{} & \negthinspace{}\negthinspace{}$\mathrm{f}_{x}$\negthinspace{}\negthinspace{} & $\frac{xy^{6}+2y^{5}-(x^{3}+2)y^{4}-2(x-2)y^{3}-(2x^{3}+7x+4)y^{2}-4xy+3x^{2}}{(y-x)(y+x)(y-1)(y+1)(y^{2}+2)}_{\,}$\rule[-0.7em]{0em}{2.1em}\tabularnewline
\hline 
\negthinspace{}\negthinspace{}$\mathrm{f}_{y}$\negthinspace{}\negthinspace{} & \negthinspace{}\negthinspace{}\negthinspace{}$\mathrm{e}_{x}$\negthinspace{}\negthinspace{}\negthinspace{}\negthinspace{} & $\frac{xy^{6}+2y^{5}-(x^{3}+2)y^{4}-(2x-4)y^{3}-(2x^{3}+7x+4)y^{2}-4xy+3x^{2}}{(y-x)(y+x)(y-1)(y+1)(y^{2}+2)}_{\,}$\rule[-0.7em]{0em}{2.1em}\tabularnewline
\hline 
 &  & \vspace{-1.1em}
\tabularnewline
\hline 
\negthinspace{}\negthinspace{}\negthinspace{}$\mathrm{e}_{x}$\negthinspace{}\negthinspace{}\negthinspace{} & \negthinspace{}\negthinspace{}$\mathrm{f}_{y}$\negthinspace{}\negthinspace{} & $\frac{y^{3}+3}{y^{2}+2}x+\frac{2y^{2}}{(x+y)(y-1)(y+1)}+\frac{2y}{(x-y)(y-1)(y+1)}_{\,}$\rule[-0.7em]{0em}{2.1em}\tabularnewline
\hline 
\negthinspace{}\negthinspace{}\negthinspace{}$\mathrm{e}_{x}$\negthinspace{}\negthinspace{}\negthinspace{} & \negthinspace{}\negthinspace{}\negthinspace{}$\mathrm{e}_{y}$\negthinspace{}\negthinspace{}\negthinspace{} & $x+\frac{x}{y^{2}+2}+\frac{2}{x+y}+\frac{1}{xy-x+y^{2}-y}+\frac{1}{xy+x+y^{2}+y}+\frac{1}{xy-x-y^{2}+y}+\frac{1}{xy+x-y^{2}-y}_{\,}$\rule[-0.7em]{0em}{2.1em}\tabularnewline
\hline 
\negthinspace{}\negthinspace{}\negthinspace{}$\mathrm{e}_{y}$\negthinspace{}\negthinspace{}\negthinspace{} & \negthinspace{}\negthinspace{}$\mathrm{f}_{x}$\negthinspace{}\negthinspace{} & $\begin{cases}
1-\frac{y+8}{6(y-1)^{2}}-\frac{11y+8}{6(y+1)^{2}}+\frac{1}{3(y^{2}+2)}, & \mathrm{if}\: x=-1,\\
-1+\frac{1}{(y-1)^{2}}-\frac{2y+1}{(y+1)^{2}}-\frac{1}{y^{2}+2}, & \mathrm{if}\: x=1,\\
x+\frac{x}{y^{2}+2}+\frac{2x}{(x-1)(x+1)(x-y)}-\frac{2x^{2}}{(x-1)(x+1)(y+x)}+\frac{2x}{(x-1)(x+1)(y-1)}-\frac{2}{(x-1)(x+1)(y+1)}, & \mathrm{otherwise}
\end{cases}$\rule[-2.2em]{0em}{5.2em}\tabularnewline
\hline 
\negthinspace{}\negthinspace{}\negthinspace{}$\mathrm{e}_{y}$\negthinspace{}\negthinspace{}\negthinspace{} & \negthinspace{}\negthinspace{}$\mathrm{e}_{x}$\negthinspace{}\negthinspace{}\negthinspace{} & $\!\negthinspace\begin{cases}
1-\frac{y+8}{6(y-1)^{2}}-\frac{11y+8}{6(y+1)^{2}}+\frac{1}{3(y^{2}+2)}, & \!\negthinspace\mathrm{if}\: x=-1,\!\negthinspace\\
-1+\frac{1}{(y-1)^{2}}-\frac{2y+1}{(y+1)^{2}}-\frac{1}{y^{2}+2}, & \mathrm{\!\negthinspace if}\: x=1,\\
\!\cdot\!\!\cdot\!\negthinspace\cdot\!+\!\frac{2}{y+x}\!-\!\frac{1}{xy+y+x^{2}+x}\!+\!\frac{1}{xy-y+x^{2}-x}\!-\!\frac{1}{xy+y-x^{2}-x}\!-\!\frac{1}{xy-y-x^{2}+x}\!+\!\cdot\!\!\cdot\!\negthinspace\cdot\!-\!\frac{1}{xy-y+x-1}\!+\!\frac{1}{xy-y-x+1}, & \mathrm{\!\negthinspace otherwise}\!\negthinspace
\end{cases}$\rule[-2.2em]{0em}{5.1em}\tabularnewline
\hline 
\end{tabular}
\end{table}

\subsection{Series and other approximations}

The discussion so far has been about transformations to alternatives
that are \textsl{equivalent} to the input wherever the input is defined
-- except perhaps approximating exact numbers in the input with approximate
Floats or approximating Floats in the input with nearby rational numbers.
This sub-section instead addresses the equally important alternatives
of transformations that \textsl{approximate} the input with \textsl{simpler
expressions}.

Closed-form exact results aren't always obtainable. Even when they
are, the results might be too bulky to convey needed insight or to
permit fast enough well-conditioned evaluation for numerous floating-point
values of the variables therein. Therefore, various kinds of approximation
are useful transformations. Also it is important to realize that the
ultimate destiny of many exact expressions is to substitute Floats
into them, in which case the resulting rounding errors might exceed
those caused by an approximate expression. Here are some examples
of appropriate optional approximate transformations for a wizard to
offer:\vspace{-0.3em}

\begin{itemize}
\item Quadrature can often be used to determine a single approximate number
for a definite integral.\vspace{-0.3em}

\item Approximate equation solving is often preferable even when compact
explicit exact solutions are obtainable.\vspace{-0.3em}

\item Generalized infinite or truncated Laurent-Puiseux series (allowing,
for example, logarithmic factors) can concisely approximate lengthy
expressions. The wizard can initialize expansion points to ones that
are most likely of interest, such as 0, infinities, and poles. If
selected, the user can adjust the expansion points and the requested
order.\vspace{-0.3em}

\item Padé approximations often have a larger region of convergence and
greater computational efficiency than power series.\vspace{-0.3em}

\item Truncated Fourier or wavelet series are often more appropriate than
expansions about a point.\vspace{-0.3em}

\end{itemize}
A well-chosen approximation can be simpler and better conditioned
than any obtainable exact result while retaining all of the important
qualitative characteristics of an exact result.

\subsection{Generating the list of generalized variables}

If a variable $v$ in the framed subexpression has an assigned value,
then it would be misleading to list that irrelevant $v$. However,
we do want to consider listing some or all of the generalized variables
in the \textsl{assigned value}, if any. We can use default simplification
for this purpose, because it replaces all assigned variables with
their values. 

On many systems \textsl{default} simplification can easily produce
results containing generalized variables that candid simplification
would eliminate. For example, the default simplification of most systems
merely reorders the factors and/or terms in the input $\left((x+1)x-x^{2}-x+2\right)/\left(y^{2}-1\right)$,
but all of the factoring and expansion transformations described in
sub-subsections \ref{sub:UnivariateFactorizationAndExpaqnsion} and
\ref{sub:MultivariateFactorExpand} transform the expression to $2/(y^{2}-1)^{2}$
or $2/\left((y-1)(y+1\right)$ or $1/(y-1)+1/(y+1)$, which all depend
on $y$ alone.

It is helpful for the wizard to recognize such \textsl{superfluous}
generalized variables. For polynomial expressions, expansion to recursive
form always eliminates superfluous variables. For other rational expressions,
reduction over a common denominator always does so.%
\footnote{If this reduction produces 0/0, then the expression is undefined for
all values of its variables, so the one and only offered alternative
can be the result of ``0/0''.%
} Thus the first thing that the wizard can do is compute such a form
to identify generalized variables that thereby disappear. One of these
transformations can be initially checked to help encourage the user
to eliminate superfluous generalized variables. However, the user
might prefer to eliminate them in a way that alters cherished structure
less drastically. To do this, the wizard could also offer the alternative,
for example, ``merely eliminate superfluous $x$ and $\ln(y)$''.
This can be accomplished by substituting simple exact constants such
as 0, 1 or -1 for those generalized variables, then applying default
simplification. If the substitution value is at a removable singularity
and thereby causes division by 0, then the wizard can backtrack and
try another simple constant. 

A generalized variable wouldn't be offered if doesn't affect ordering
and no optional transformation is applicable to it. For example, it
is pointless to list $x$ if it only occurs as the argument in $\sin\left(x\right)$.
However, it might be worthwhile to list the generalized variable $\sin\left(x\right)$,
depending on how it occurs in the default-simplified result. As another
example, \textsl{none} of the transformations or ordering choices
being discussed here are applicable to the subexpression $3x^{2}$.

There will usually be an initial ``Choose main variable'' dialog
if there is more than one applicable generalized variable. If so,
and after choosing the transformation with respect to the selected
main variable there is still more than one applicable generalized
variable for some maximal subexpression that doesn't contain the main
variable, then there will be another dialog labeled instead ``Choose
next most main variable'', and so on until the user aborts the investigation
or accepts replacements.

\subsection{Estimating the number of common transformations for each generalized
variable choice}

Quick syntactic checks can identify some opportunities for transforming
a framed subexpression with respect to a generalized variable. For
example,\vspace{-0.3em}

\begin{itemize}
\item Expansion is applicable to any variable that occurs in a multinomial
raised to an integer power or a multinomial multiplied by any subexpression.\vspace{-0.3em}

\item A common denominator is applicable to any variable that occurs to
a negative power in a sum.\vspace{-0.3em}

\item Angle sum expansion is applicable if the argument of a sinusoid is
a sum.\vspace{-0.3em}

\item The transformations $\sin(u)^{2}\rightarrow1-\cos(u)^{2}$ or $\sin(u)^{n}\rightarrow(\sin(nu)+\cdots)/2^{n}$
are applicable if $\sin(u)$ is raised to any integer power exceeding
1.\vspace{-0.3em}

\end{itemize}
Quick syntactic checks can also \textsl{preclude} some opportunities
for transforming a framed subexpression with respect to a generalized
variable. For example, expansion is precluded if the framed subexpression
is monomial or linear in the variable.

Another ordering heuristic is that the number of applicable transformations
is likely to increase with the degree of a variable in a numerator
or denominator, and more transformations are associated with high
degree denominators than high degree numerators, because denominator
factorizations can also enable partial fraction expansions.

If time remains in the 0.1 second acceptable delay for generating
and displaying the first dialog box, then we can compute more accurate
estimates for the number of common applicable transformations for
each variable: For the rational aspects of a framed subexpression,
minimal-effort reduction over a common denominator reveals the numerator
and denominator degrees of every top-level generalized variable, which
are all of those that can be affected by top-level factoring or expansion.
Minimal effort here means using partially factored form -- preferably
recursive -- as described in \cite{Stoutemyer10commandments}. If
this reduced partially factored form differs from the framed subexpression,
then it is already one applicable transformation. Moreover, expansion
to a polynomial plus a proper fraction is applicable to every such
generalized variable whose degree in the resulting numerator is at
least as large as the degree in the denominator; and it is easy to
compute these degrees for partially factored forms.

If time still remains in the 0.1 second acceptable delay, then we
can compute more accurate estimates by factoring the common denominator
with respect to all of its generalized variables. From this it is
easy to determine which successively lesser factorizations would combine
two or more factors containing that main variable and thus determine
the number of distinct named partial fraction expansions with respect
to that variable and a lower bound on the number of distinct named
factorizations over a common denominator.

If time still remains, then the wizard can factor the numerator to
better estimate the number of named factorizations over a common denominator.

To reduce the chance of exceeding 0.1 second before \textsl{any} factorization
is achieved, it could be done in levels, such as term content with
respect to every variable, then square free, etc. through, say, exact
reasonably absolute factorization followed by absolute factoring over
the complex Floats. If the allotted 0.1 seconds runs out before completing
this agenda, then we simply use the best estimates that we have at
that time. Moreover, we can continue to refine and update the estimates
after the initial display, without changing the initial order of the
generalized variable buttons.

\subsection{Recognizing applicable transformations}

The exact and approximate factorizations with respect to all variables
is a useful point of departure for computing alternative results regardless
of what variable the user chooses. Therefore it is helpful to proceed
computing those factorizations in the background while users ponder
their choice of variable.

After a user chooses a variable, the wizard can complete the distinct
factorization levels by appropriately expanding pairs of factors,
which is fast. To convey the strongest possible information, when
labeling the displayed alternatives or partially elided versions thereof,
they are labeled with the most thorough applicable factoring level.
For example, if the only distinct factorizations are factorizations
over the Gaussian integers and over the integers, then the later would
not be labeled ``square-free'' even though it is that as well. If
there is only one factored alternative, then it could be labeled merely
``factored'' for brevity, but with a more detailed phrase displayed
upon mouse-over. This provides an unintimidating learning opportunity
for mathematically unsophisticated users.

If there is a denominator and it contains the chosen variable, the
wizard can then proceed to compute the polynomial part and proper
fraction, followed by any distinct partial fraction levels with respect
to that variable.

\section{\noindent \label{sec:Summary}If I want this interface, why haven't
I implemented it?}

Good question. The reasons are:\vspace{-0.3em}

\begin{itemize}
\item I am not an interface programmer.\vspace{-0.3em}

\item It is best done by a team.\vspace{-0.3em}

\item Some aspects will require access to proprietary internals that are
inaccessible to outsiders for systems that aren't open source.\vspace{-0.3em}

\item It will require testing feedback from numerous users. Surely some
of the ideas presented here won't work out well in practice, and better
ideas will occur.\vspace{-0.3em}

\item If it isn't the default interface distributed with a computer algebra
system, then it is unlikely to be known to most users, and the system
is likely to evolve in a way so that the alternative interface no
longer works.\vspace{-0.3em}

\item For various reasons, corporations are often unwilling to adopt and
maintain packages written by outsiders as first class parts of their
system -- thoroughly and seamlessly integrated into the system and
the documentation with no need for building or loading by users.\vspace{-0.3em}

\end{itemize}
Therefore, although I would be delighted to help, the purpose of this
manifesto is to stimulate computer algebra users to request better
interfaces and stimulate decision makers to build them.

\noindent \begin{flushright}
``\textsl{If you build it, they will come}.''\\
-- an apt misquote from \textsl{Field of Dreams}\vspace{1em}

\par\end{flushright}

Computer algebra users of the world: the squeaky wheels get the grease!

\section*{Acknowledgments}

I thank Bill Gosper for information about Lisp Machine Macsyma and
Norbert Kajler for helpful suggestions. Jacques Carette provided so
many extraordinarily good suggestions that he should be a co-author
-- except for the conflict of interest that he is an unmasked referee
for the forthcoming primary version of this article in the \textsl{ACM
Communications in Computer Algebra}.

\section*{Appendix: More transformations for rational expressions}

It is worthwhile to list in one place most of the many known transformations
that might be of general interest. Subsection \ref{sub:CommonRationalTransformations}
discussed factoring, common denominators and expansion. Here are some
additional transformations for the rational aspect of expressions:

\subsection*{A.1: Basic -- of interest to most users}

The wizard should automatically try the following transformations
in the background because they can dramatically decrease the bulk
of an expression and reveal important structure:

\subsubsection*{\noindent \label{sec:Polynomial-shifts}Polynomial shifts}

\noindent Most of the factorization and expansion levels leave polynomial
sub-expressions that often have more than two terms. Sometimes merely
re-expressing such a multinomial in terms of optimally shifted variables
can greatly reduce the number of terms and/or the size of coefficients.
For example,
\begin{eqnarray*}
\left(y^{2}\negthinspace-\negthinspace4y\negthinspace+\negthinspace4\right)x^{3}+\left(3y^{2}\negthinspace-\negthinspace12y\negthinspace+\negthinspace12\right)x^{2}+\left(3y^{2}\negthinspace-\negthinspace12y\negthinspace+\negthinspace12\right)x+y^{2}-4y+1 & \boldsymbol{\rightarrow} & \left(y\negthinspace-\negthinspace2\right)^{2}\negthinspace\left(x\negthinspace+1\right)^{3}+8,
\end{eqnarray*}
\[
531441\, x^{6}+2834352\text{\,}x^{5}+6298560\, x^{4}+7464960\, x^{3}+4976640\text{\,}x^{2}+1769472\, x+262151\rightarrow\left(9x+8\right)^{6}+7\,.
\]
Articles \cite{Giesbrecht et al,Grigoriev and Lakshman}, give algorithms
for computing optimal shifts. As a quick preclusion test, it is not
worth shifting a polynomial with respect to a variable if the polynomial
is monomial or binomial with respect to that variable.

\subsubsection*{\noindent Polynomial decomposition}

\noindent Complementary to such shift decompositions, Kozen and Landau
\cite{Kozen and Landau} describe an efficient algorithm for completely
decomposing a univariate polynomial $p(x)$ into nested polynomials
\[
p_{1}\left(p_{2}\left(\ldots\left(p_{m}\left(x\right)\right)\ldots\right)\right)
\]
with each $p_{k}\left(t\right)$ of degree at least 2 in \textit{t}.
For example, the irreducible polynomial

\noindent 
\begin{eqnarray*}
P(x) & := & x^{12}+4x^{10}+x^{9}+6x^{8}+3x^{7}+4x^{6}+3x^{5}+x^{4}+x^{2}+7\\
 & \to & \left(x^{3}+x\right)^{4}+\left(x^{3}+x\right)^{3}+7.
\end{eqnarray*}
Their algorithm also applies recursively to multivariate polynomials
represented recursively. As a quick preclusion test, such decompositions
are inapplicable with respect to a variable of degree less than 4
or having few terms. There are also algorithms for other kinds of
univariate and multivariate polynomial decompositions, as described,
for example, in \cite{FaugereAndPerret,vonZerGathenEtAl,vonZurGathenMultivariate,vonZurGathenWild,WattDecompSymbolic}.

Polynomials rewritten in these ways can reveal significant structure,
help precondition an expression for efficient repeated numeric evaluation
or reduced rounding error, and facilitate solutions of higher-degree
polynomial equations or systems of equations. For example, with the
above decomposition, masochists could apply the quartic formula then
the cubic formula to express the zeros of \textit{P}(\textit{x}) in
terms of radicals.

\subsubsection*{Linear combinations of powers}

At least since Pythagoras, people have been interested in representing
numbers and non-numeric expressions as sums, differences, or general
linear combinations of powers of other expressions. For example, if
an expression can be rewritten as a sum of even powers of real subexpressions,
then the expression is thereby proven to be nonnegative for all real
values of these subexpressions. Algorithms for such transformations
can be found in, for example, \cite{PowersAndWoermann,Reznick,VandenbergheAndBoyd}.

\subsection*{A.2: Advanced -- of interest only to experts}

Here are some transformations that should be tried only in response
to the \fbox{\sf{Advanced}} button because they would probably intimidate
and distract most users without any compensating benefit to them.

\subsubsection*{Expression in terms of orthogonal polynomials}

Important optional transformations include a change of basis from
monomials to orthogonal polynomials, which can\vspace{-0.3em}

\begin{itemize}
\item yield more concise results,\vspace{-0.3em}

\item yield results less subject to magnified rounding errors,\vspace{-0.3em}

\item reveal patterns that otherwise wouldn't be apparent, or\vspace{-0.3em}

\item suggest efficient accurate min-max truncated approximations.\vspace{-0.3em}

\end{itemize}
Many computer algebra systems provide functions that return one of
various classic orthogonal polynomials of a specified degree in a
specified variable using the monomial basis. However, most of the
systems provide little or no automation for:\vspace{-0.3em}

\begin{itemize}
\item converting ordinary polynomials to linear combinations of orthogonal
polynomials specified by \textsl{symbols} such as the Chebyshev polynomials
of the first kind $T_{0}(z)$, $T_{1}(z)$, $\ldots$ ;\vspace{-0.3em}

\item propagating linear combinations of such polynomials into other linear
combinations thereof \textsl{exactly} through rules such as
\begin{eqnarray*}
T_{m}(z)\, T_{n}(z) & \rightarrow & \frac{1}{2}\, T_{\left|n-m\right|}(z)+\frac{1}{2}\, T_{m+n}(z),\\
T_{m}\left(T_{n}(z)\right) & \rightarrow & T_{mn}(z),\\
\int T_{n}(z)\, dz & \rightarrow & \begin{cases}
\frac{1}{4}\left(T_{0}(x)+T_{2}(x)\right), & \mathrm{if}\: n=1,\\
\frac{1}{2-2n}\, T_{n-1}(z)+\frac{1}{2+2n}\, T_{n+1}(z), & \mathrm{otherwise},
\end{cases}
\end{eqnarray*}
 and \textsl{approximately} through infinite or truncated series expansions;\vspace{-0.3em}

\item efficiently and accurately substituting Floats into expressions involving
combinations of such functions: Rather than substituting a number
$z_{0}$ for $z$ in the monomial-basis representations of the symbols
$T_{0}(z)$, $T_{1}(z)$, $\ldots$ in a combination thereof, it is
much faster and more accurate to compute the successive $T_{k}(z_{0})$
from the recurrence $T_{n}(z_{0})\leftarrow2z_{0}\, T_{n-1}(z_{0})-T_{n-2}(z_{0})$.\vspace{-0.3em}

\end{itemize}
Of the various named orthogonal polynomials, $T_{0}(z)$, $T_{1}(z)$,
$\ldots$ are probably most important. Therefore, Trefethen and others
\cite{TrefethenEtAl;} developed a powerful \textsc{Matlab} package
that automates effective use of these polynomials for many applications,
using IEEE double Floats to represent the coefficients. Fateman \cite{FatemanChebyshev}
implemented some of these capabilities in Maxima to take advantage
of its variable precision floating point and exact rational arithmetic.
Such transformations and analogous ones for other orthogonal polynomials
would be a welcome addition to many computer algebra systems.

\subsubsection*{Expression in terms of symmetric polynomials}

Analogous transformations for the most common kinds of \textsl{symmetric
polynomials} would be another welcome addition. Such polynomials can
help reduce the curse of dimensionality for problems that exhibit
symmetries when pairs of variables are interchanged. As a start toward
this, the \textsl{Mathematica} function $\mathrm{SymmetricReduction}\,[expression,\,\{v_{1},v_{2},\ldots v_{n}\},\,\{s_{1},s_{2,},\ldots,s_{n}\}]$
returns the pair $\{p,r\}$ where $\mathit{expression}$ depends on
variables $\{v_{1},v_{2},\ldots v_{n}\}$, $p$ is the symmetric component
of $\mathit{expression}$ in terms of symbols $\{s_{1},s_{2,},\ldots,s_{n}\}$
representing the \textsl{elementary symmetric polynomials} through
degree $n$, and $r$ is the residual of $\mathit{expression}$ that
can't be so represented. To convert a symmetric result back to the
original variables, function $\mathrm{SymmetricPolynomial[}m,\{v_{1},v_{2},\ldots v_{n}\}]$,
returns the $m$\textsuperscript{th} elementary symmetric polynomial
using variables $\{v_{1},v_{2},\ldots v_{n}\}$. Sturmfels \cite{Sturmfels}
contains algorithms for transforming to and from symmetric polynomials.

\subsubsection*{\noindent Rational Decomposition}

\noindent We can attempt separate polynomial decompositions on every
numerator and denominator polynomial in an expression. However, there
are also algorithms for decomposing \textsl{ratios} of polynomials
into \textsl{nested ratios}, as discussed in \cite{Ayad and Fleischmann,Gutierrez et al,Zippel}.
As an example from the first of these articles, but using recursive
form and primitive normalization, the ratio
\begin{multline*}
\dfrac{\left(y^{2}+2z^{2}y+z^{4}-81\right)x^{2}-2y\cdot\left(y^{5}+z^{2}y^{4}+225\, z\right)x+y^{2}\left(y^{8}-625\, z^{2}\right)}{\left(y^{2}+2z^{2}y+z^{4}-162\right)x^{2}-2y\cdot\left(y^{5}+z^{2}y^{4}+450\, z\right)x+y^{2}\left(y^{8}-1250\, z^{2}\right)}\\
\rightarrow\begin{cases}
1, & \:\mbox{if}\;9x+25zy=0,\\
\dfrac{\left(\dfrac{\left(y+z^{2}\right)x-y^{5}}{9x+25\, zy}\right)^{2}-1}{\left(\dfrac{\left(y+z^{2}\right)x-y^{5}}{9x+25\, zy}\right)^{2}-2}, & \:\mbox{otherwise}.
\end{cases}
\end{multline*}

This example also illustrates that rational decomposition can introduce
new removable singularities in the nested form, such as on the manifold
$9x+25zy=0$ for this example. We can avoid this by clearing the nested
denominators but preserving the nested polynomial components thereof
to obtain \textsl{correlated} polynomial decompositions of the numerator
and denominator:
\begin{eqnarray*}
\dfrac{\left(\dfrac{\left(y+z^{2}\right)x-y^{5}}{9x+25zy}\right)^{2}-1}{\left(\dfrac{\left(y+z^{2}\right)x-y^{5}}{9x+25zy}\right)^{2}-2} & \rightarrow & \frac{\left(\left(y+z^{2}\right)x-y^{5}\right)^{2}-\left(9x+25\, zy\right)^{2}}{\left(\left(y+z^{2}\right)x-y^{5}\right)^{2}-2\left(9x+25\, zy\right)^{2}}.
\end{eqnarray*}

If desired, for this example we can then further factor the difference
in two squares in the numerator and in the denominator to obtain the
factorization over $Z[\sqrt{2}\,]$.

\noindent 
\[
\frac{\left(\left(y+z^{2}+9\right)x-y^{5}+25\, zy\right)\left(\left(y+z^{2}-9\right)x-y^{5}-25\, zy\right)}{\left(\left(y+z^{2}+9\sqrt{2}\right)x-y^{5}+25\sqrt{2}\, zy\right)\left(\left(y+z^{2}-9\sqrt{2}\right)x-y^{5}-25\sqrt{2}\, zy\right)}.
\]

\noindent The numerator factorization could easily have been computed
from the original numerator. However, the required $\sqrt{2}$ algebraic
extension necessary to factor the multivariate denominator would be
more difficult to determine without the intervening rational decomposition.

\subsubsection*{\noindent Continued fractions}

\noindent Continued fractions are another type of compound-fraction
representation for rational expressions. Acton \cite{Acton} lists
three different variants of continued fractions together with algorithms
for converting between them and a reduced ratio. Cuyt and Verdonk
\cite{Cuyt and Verdonk} review methods for multivariate continued
fractions. As an example of a continued fraction expansion that reveals
a simple pattern:
\[
\frac{\left(z^{4}-105\, z^{2}+945\right)z}{15\left(z^{4}-28\, z^{2}+63\right)}\:|\: z^{2}\notin\left\{ 63,\,\dfrac{45}{2},\,\dfrac{3}{2}\left(35\pm\sqrt{805}\right)\right\} \to\cfrac{z}{1-\cfrac{z^{2}}{3-\cfrac{z^{2}}{5-\cfrac{z^{2}}{7-\cfrac{z^{2}}{9}}}}}.
\]

Here a constraint was appended to the input to avoid the \textsl{appearance}
of contracting the domain of definition because of removable singularities
introduced by the continued fraction. If instead we used a piecewise
result, then it would have 9 pieces, 8 of which are constants that
can be determined by substituting the two square roots of each element
in the constraint set into the original expression.

Actually, if the computer algebra system automatically handles unsigned
zeros and infinities correctly, then with exact computation the continued
fraction form evaluates to the correct finite values even at these
removable singularities. However, unlike the reduced ratio, the continued
fraction form \textsl{might} be less accurate due to catastrophic
cancellation near those introduced singularities. Therefore it is
worth alerting the user to these singularities by either appending
an input constraint or producing a piecewise result.%
\footnote{With correct handling of infinities, a continued fraction can be defined
at infinity where a reduced common denominator is \textsl{not}. For
example, $1/(1+1/z)\rightarrow1$ at $z=\infty$, where the corresponding
reduced ratio $z/(z+1)\rightarrow\infty/\infty$ is indeterminate.
Most people would prefer having a result defined at all \textsl{finite}
values of variables to being defined at \textsl{infinite} values,
but a mere division can turn 0 into an infinite value.%
}

\subsubsection*{\noindent Hornerized forms}

\noindent In its simplest form, Horner's rule is the factoring out
of the least power of a variable from successively lower-degree terms.
For example,
\begin{multline*}
-1234321\, x^{7}-1234321\, x^{5}+2468642\, x^{4}+7x^{3}+14x-21\\
\rightarrow((((-1234321\, x^{2}-1234321)\, x+2458642)\, x+7)\, x^{2}+14)\, x-21.
\end{multline*}

It is also worth partially factoring out units and numeric content
to the extent that it reduces bulk or the number of operations. For
example,
\begin{multline*}
-1234321\, x^{7}-1234321\, x^{5}+2468642\, x^{4}+7x^{3}+14x-21\\
\rightarrow((-1234321((x^{2}+1)\, x-2)\, x+7)\, x^{2}+14)\, x-21.
\end{multline*}

Horner's rule often leads to faster evaluation when substituting numbers
for variables, which is particularly important in situations such
as plotting, where substitution is done many times for different values.
Horner's rule also often improves accuracy for approximate arithmetic,
because the operands of a catastrophic cancellation are closer to
the input numbers, hence less contaminated with rounding error.

Horner's rule can be viewed as factoring out term content term by
term, starting with the highest-degree terms at each level. With this
viewpoint, we can apply it to multinomials throughout an expression
in all of the above special forms. Ceberio and Kreinovich \cite{Ceberio and Kreinovich}
discuss greedy algorithms for computing efficient multivariate Hornerized
forms.

Another transformation that can enable faster evaluation when substituting
numbers for variables is to factor out an integer common divisor of
the \textsl{exponents} from a product of powers. For example, if the
powers are done with the help of repeated squaring, then
\[
(y+3)^{6}x^{4}\rightarrow\left(\left(y+3\right)^{3}x^{2}\right)^{2}
\]
uses only five multiplications rather than six.%
\footnote{However, for most systems default simplification automatically distributes
the outer exponent 2 over the two factors unless something special
is done to prevent it.%
}

\end{document}